\documentclass[aps,prl,amsmath,amssymb,reprint,floatfix,superscriptaddress,nofootinbib]{revtex4-2}
\usepackage{graphicx}
\usepackage{bm}
\usepackage{hyperref}
\usepackage{float}
\hypersetup{
    colorlinks=true,
    linkcolor=blue,
    citecolor=blue,
    urlcolor=blue
}

\begin{document}

\title{Enhanced Parallel Quantum Sensing via Intensity-Tailored Optical Addressing}

\author{Yu-Xuan He}
\thanks{These authors contributed equally to this work.}
\affiliation{Center for Quantum Information, Institute for Interdisciplinary Information Sciences, Tsinghua University, Beijing 100084, People's Republic of China}
\author{Ya-Nan Lu}
\thanks{These authors contributed equally to this work.}
\affiliation{Center for Quantum Information, Institute for Interdisciplinary Information Sciences, Tsinghua University, Beijing 100084, People's Republic of China}
\affiliation{Hefei National Laboratory, Hefei 230088, People's Republic of China}

\author{Zhi-Yun Zheng}
\affiliation{Center for Quantum Information, Institute for Interdisciplinary Information Sciences, Tsinghua University, Beijing 100084, People's Republic of China}

\author{Jia-Hao Tian}
\affiliation{Center for Quantum Information, Institute for Interdisciplinary Information Sciences, Tsinghua University, Beijing 100084, People's Republic of China}

\author{Yan-Qing Liu}
\affiliation{Center for Quantum Information, Institute for Interdisciplinary Information Sciences, Tsinghua University, Beijing 100084, People's Republic of China}
\affiliation{Hefei National Laboratory, Hefei 230088, People's Republic of China}

\author{Bin-Xiang Qi}
\affiliation{Center for Quantum Information, Institute for Interdisciplinary Information Sciences, Tsinghua University, Beijing 100084, People's Republic of China}
\affiliation{Hefei National Laboratory, Hefei 230088, People's Republic of China}

\author{Dong-Ling Deng}
\email{dldeng@tsinghua.edu.cn}
\affiliation{Center for Quantum Information, Institute for Interdisciplinary Information Sciences, Tsinghua University, Beijing 100084, People's Republic of China}
\affiliation{Hefei National Laboratory, Hefei 230088, People's Republic of China}
\affiliation{Shanghai Qi Zhi Institute, Shanghai 200232, People's Republic of China}

\author{Lu-Ming Duan}
\email{lmduan@tsinghua.edu.cn}
\affiliation{Center for Quantum Information, Institute for Interdisciplinary Information Sciences, Tsinghua University, Beijing 100084, People's Republic of China}
\affiliation{Hefei National Laboratory, Hefei 230088, People's Republic of China}
\affiliation{New Cornerstone Science Laboratory, Beijing 100084, People's Republic of China}

\author{Pan-Yu Hou}
\email{houpanyu@tsinghua.edu.cn}
\affiliation{Center for Quantum Information, Institute for Interdisciplinary Information Sciences, Tsinghua University, Beijing 100084, People's Republic of China}
\affiliation{Hefei National Laboratory, Hefei 230088, People's Republic of China}
\date{\today}

\begin{abstract}

Parallel quantum sensing enables simultaneous measurements at spatially distributed locations, but heterogeneous optical responses and limited excitation power constrain the number and precision of accessible spatial comparisons.
Here we demonstrate parallel quantum sensing with nitrogen-vacancy centers in nanodiamonds by tailoring the position and intensity of each excitation beam.
Using a spatial light modulator, we selectively illuminate individual nanodiamonds and adapt their excitation intensities through fluorescence feedback to compensate for variations in their optical responses.
For 139 spatially distributed nanodiamonds, fluorescence feedback reduces the relative root-mean-square variation in fluorescence from 97.6\% under uniform-weight addressing to 1.3\% under tailored addressing.
Compared with wide-field illumination, tailored addressing yields more than three times as many nanodiamonds with resolvable electron-spin resonances and ten times as many sensor pairs for spatial comparisons.
The median sensitivity to pairwise resonance-frequency differences improves by a factor of 3.5, and the total sensitivity to a common resonance-frequency shift improves by a factor of 3.7.
These results establish tailored optical addressing as an effective approach for parallel quantum sensing with heterogeneous sensors, with potential applications in multiplexed intracellular thermometry and spatially resolved sensing in complex environments.
\end{abstract}

\maketitle

\noindent

\textit{Introduction}---The nitrogen-vacancy (NV) centers in diamond have emerged as versatile quantum sensors for magnetic fields and temperature~\cite{Taylor2008magnetometer,Maze2008magnetic,Balasubramanian2008imaging,Maletinsky2012scanning,Grinolds2013singleSpin,Acosta2010temperature,Neumann2013thermometry,Toyli2013thermometry,Kucsko2013thermometry,Barry2020sensitivity,LeSage2013Optical}.
When hosted in nanodiamonds (NDs), they serve as localized probes in complex biological environments, including living cells~\cite{Kucsko2013thermometry,LeSage2013Optical,McGuinness2011tracking,Fujiwara2020real,Simpson2017neuronal,Sotoma2021insitu,Nie2021Quantum,Sigaeva2022Diamond,Consiglio2026Optically,Miller2020spin,Steinert2013Magnetic}.
In such environments, physical quantities can vary across space and evolve over time~\cite{Fujiwara2020real,Nishimura2021wide,Sow2020high}.
For example, recent works have reported temperature differences between intracellular locations~\cite{Lee2025Organelle}, and temporal fluctuations in local electromagnetic noise associated with cellular activity~\cite{Xu2025Multiphysiology}.
Resolving these spatiotemporal variations calls for parallel optical readout with sufficient measurement precision and temporal resolution.

Conventional confocal microscopy provides localized excitation and detection, but sequential scanning limits the temporal resolution of measurements at multiple locations~\cite{Nishimura2021wide,Scholten2021widefield}.
Wide-field imaging enables simultaneous fluorescence detection from multiple NDs~\cite{Nishimura2021wide}.
However, wide-field illumination also excites regions without target sensors, potentially increasing background fluorescence, such as cellular autofluorescence~\cite{Igarashi2012real,Hui2014wide}. Such nonselective illumination is also undesirable in biological measurements, where phototoxicity can limit the permissible light dose~\cite{Icha2017phototoxicity,Laissue2017phototoxicity}.
Moreover, spatial variations in excitation intensity and differences in the number of NV centers per nanodiamond give rise to heterogeneous fluorescence signals~\cite{Sow2020high,Reineck2017Bright,Oshimi2024bright}.
Under limited optical power, weakly emitting NDs exhibit low spin-readout signal-to-noise ratios, reducing their sensing precision~\cite{Nishimura2021wide,Hopper2018spinreadout}.
This limits both the number of sensors available for spatial comparisons and the precision of these comparisons.

Spatially programmable illumination offers a route to addressing these limitations by controlling both the spatial distribution and local intensity of optical excitation~\cite{Pasienski2008holographic,Gaunt2012holography,Nogrette2014holographic,Cai2021parallel,Cai2023highthroughput}.
Spatial light modulators have been widely used to generate optical-tweezer arrays~\cite{Nogrette2014holographic,Tamura2016highly,Kim2019large,Chew2024ultraprecise}. Recent experiments have demonstrated parallel optical addressing and spin readout of more than 100 individually resolved single NV centers in diamond~\cite{Cheng2025massively,Cambria2025scalable}. Such programmability allows the available excitation power to be distributed according to the optical responses of individual sensors, providing a means to enhance parallel quantum sensing in heterogeneous systems.

\begin{figure*}[t!]
  \centering
  \includegraphics[width=\textwidth]{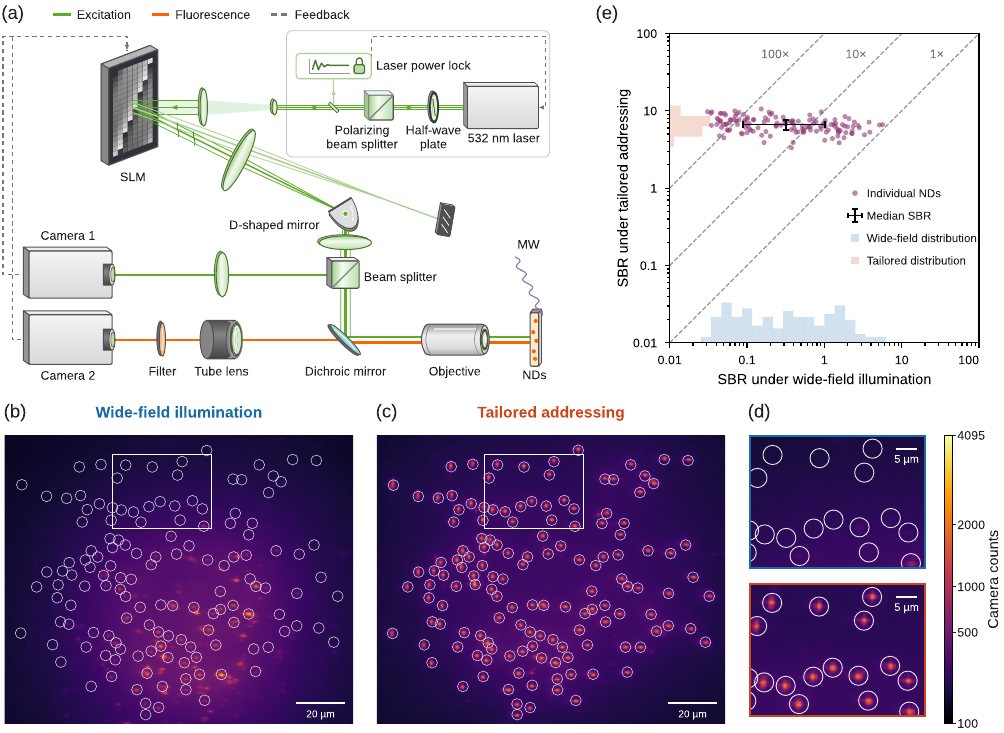}
\caption{\textbf{Intensity-tailored optical addressing of nanodiamonds.}
(a) Experimental setup.
A spatial light modulator (SLM) shapes the 532-nm power-stabilized  excitation beam into independently weighted spots at individual nanodiamond (ND) locations.
Camera~1 monitors the excitation pattern, and Camera~2 records the NV fluorescence through a long-pass filter.
Dashed lines indicate feedback paths. MW, microwave.
(b), (c) Fluorescence images of the same 139 NDs under wide-field illumination and tailored addressing, respectively. Wide-field illumination uses a Gaussian beam with a $1/e^2$ diameter of 85 $\mu$m. The camera exposure times are 32.59 ms in (b) and 43.00 ms in (c), chosen to keep the brightest ND just below camera saturation. Circles indicate the ND positions. (d) Magnified views of the regions outlined in (b) (top) and (c) (bottom). All fluorescence images share the same color scale. (e) Signal-to-background ratio (SBR) of individual NDs under tailored addressing versus wide-field illumination. Dashed lines indicate improvement factors of 1, 10, and 100. Histograms along the axes show the corresponding SBR distributions. The point marks the median SBRs, with horizontal and vertical bars spanning the 25th–75th percentiles of the respective distributions.}
\label{fig:setup}
  \label{fig:setup}
\end{figure*}

Here we demonstrate intensity-tailored optical addressing of NV centers in nanodiamonds to enhance parallel quantum sensing. Using holographic excitation and fluorescence feedback, we reduce the relative root-mean-square fluorescence deviation across 139 spatially distributed NDs from 97.6\% under uniform-weight addressing to 3.5\% within 15 s and to 1.3\% within 44 s. We then characterize the sensing performance through electron spin resonance measurements under wide-field illumination and tailored addressing. Tailored addressing increases the number of NDs with resolvable spin resonances from 43 to 135, extending the number of accessible spatial comparisons from 903 to 9045. From the measured spectra, we evaluate the sensitivity to resonance-frequency differences between pairs of NDs and to a frequency shift common to all NDs. The median pairwise sensitivity improves by a factor of 3.5, and the total sensitivity  to a common shift improves by a factor of 3.7. Together, these gains provide a route to probing spatial variations at more locations and on shorter timescales, with potential applications in multiplexed intracellular thermometry and spatially resolved sensing in complex environments.

\begin{figure*}[t!]
  \centering
  \includegraphics[width=\textwidth]{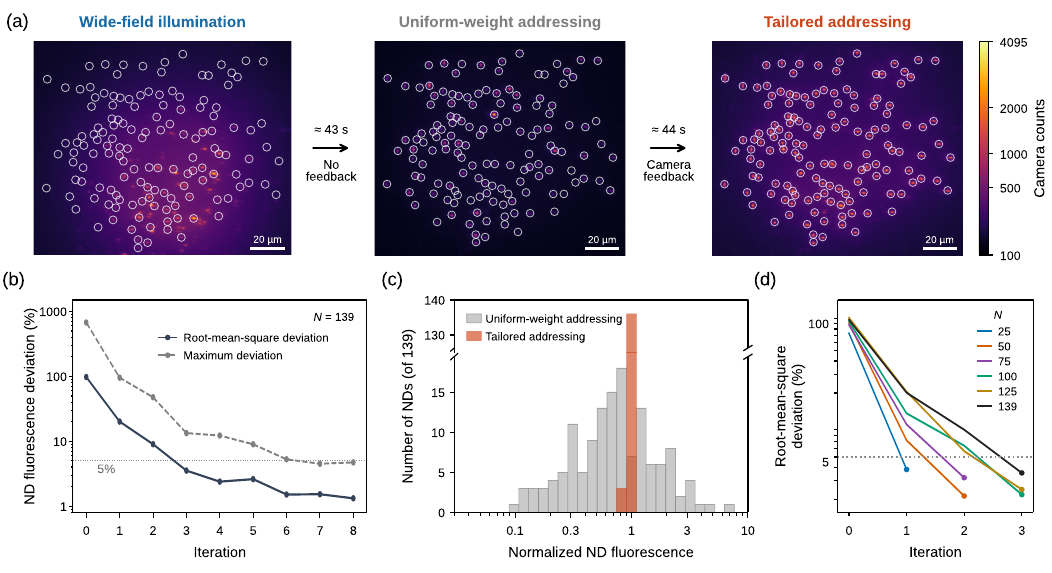}
\caption{\textbf{Fluorescence feedback optimization for intensity-tailored optical addressing.}
(a) Fluorescence images of the 139 nanodiamonds under wide-field illumination, uniform-weight addressing, and tailored addressing after eight feedback iterations.
The corresponding exposure times are 32.59, 6.99, and 43.00~ms.
Arrows indicate the elapsed times of approximately 43~s for initial pattern generation without feedback and 44~s for eight fluorescence feedback iterations.
Circles mark the ND positions, and all images share the same color scale.
(b) Root-mean-square (RMS) fluorescence deviation (solid line) and maximum deviation (dashed line) as functions of feedback iteration for 139 NDs.
Iteration 0 corresponds to uniform-weight addressing.
(c) Distributions of the ND fluorescence, normalized to the mean over all NDs under uniform-weight addressing (gray) and tailored addressing (orange).
(d) RMS fluorescence deviation versus feedback iteration for different numbers $N$ of addressed NDs, selected in descending order of their fluorescence intensity under wide-field illumination. Each curve ends when the RMS deviation falls below 5\%. In (b) and (d), dotted lines indicate the 5\% reference level.}
  \label{fig:feedback}
\end{figure*}

\textit{Intensity-Tailored Optical Addressing}---We implement intensity-tailored optical addressing using a phase-only spatial light modulator (SLM) in a wide-field microscope [Fig.~\ref{fig:setup}(a)]. The SLM shapes a power-stabilized 532-nm laser beam into multiple excitation spots with independently controlled positions and intensities.
The excitation pattern is projected onto individual NDs through a microscope objective (20×, NA = 0.42), while a reference camera (Camera 1) monitors the SLM-generated patterns for optical calibration. The NV fluorescence is collected through the same objective, separated from the excitation light by a dichroic mirror, and imaged onto a second camera (Camera 2) through a long-pass filter.
Microwave pulses control the NV electron spins, and spatially resolved fluorescence detection provides parallel spin readout. The measured fluorescence also serves as feedback for optimizing the excitation intensity of each ND.

We first compare wide-field illumination and tailored addressing using the same 139 spatially distributed NDs [Figs.~\ref{fig:setup}(b) and \ref{fig:setup}(c)]. Under wide-field illumination, the SLM generates a single Gaussian beam with a $1/e^2$ diameter of 85 $\mu$m and an optical power of 13.8 mW at the sample (Supplementary Information). The resulting fluorescence image exhibits pronounced variations in brightness among the NDs. While some NDs appear bright, many others are barely distinguishable from the background. In contrast, tailored addressing produces bright and relatively uniform fluorescence signals across nearly all NDs, including those that are difficult to resolve under wide-field illumination [Fig.~\ref{fig:setup}(d)]. For tailored addressing, approximately 6 mW of the 13.8 mW incident on the SLM reaches the sample owing to the finite diffraction efficiency of the hologram. In both cases, the camera exposure time is chosen to keep the brightest ND just below the saturation level of the 12-bit camera, yielding 32.59 ms and 43.00 ms, respectively. Thus, improved fluorescence detection is achieved despite the lower total optical power at the sample.

To quantify this improvement, we evaluate the signal-to-background ratio (SBR) of each ND, defined as the background-subtracted peak fluorescence signal divided by the mean local background, corresponding to the Weber contrast. Figure~\ref{fig:setup}(e) compares the SBR of the same 139 NDs under the two illumination schemes, with the marginal histograms showing the corresponding distributions. Under wide-field illumination, the SBR varies widely among the NDs, with 101 of the 139 NDs below unity and a median of 0.32. In contrast, tailored addressing produces a narrower distribution centered at substantially higher SBR values, with a median of 6.6. The SBR of individual NDs improves by a median factor of 22. These results demonstrate that tailored addressing enables more consistent fluorescence detection across the NDs, including those barely distinguishable from the background under wide-field illumination.

We next examine how intensity-tailored optical addressing is implemented through spatially selective excitation and fluorescence feedback (Fig.~\ref{fig:feedback}).
We first calibrate the optical system using camera-based wavefront reconstruction~\cite{Schroff2024rapid} and identify 139 spatially separated NDs from a wide-field fluorescence image.
A weighted Gerchberg--Saxton algorithm~\cite{DiLeonardo2007computer} then generates a hologram that produces Gaussian excitation spots with a $1/e^{2}$ diameter of 5~$\mu$m at the selected ND locations (Supplementary Information). Figure~\ref{fig:feedback}(a) shows the progression from wide-field illumination to uniform-weight addressing and finally tailored addressing. Despite spatially selective excitation, uniform excitation weights still produce substantial fluorescence variations owing to the heterogeneous optical responses of individual NDs.

To compensate for these variations, we iteratively adjust the excitation weights based on the measured fluorescence relative to the mean, redistributing the optical power towards NDs with weaker fluorescence while keeping the total weight constant. Details of the feedback optimization procedure are provided in the Supplementary Information. Figure~\ref{fig:feedback}(b) shows that the relative root-mean-square fluorescence deviation decreases from 97.6\% under uniform-weight addressing to 20.1\% after one iteration and 3.5\% after three iterations. After eight iterations, it reaches 1.3\%, with a maximum deviation of 4.7\%. Correspondingly, the fluorescence distribution becomes narrowly concentrated around its mean [Fig.~\ref{fig:feedback}(c)].

We further investigate the feedback speed and its dependence on the number of addressed NDs. For 139 NDs, the root-mean-square deviation falls below 5\% within three iterations, requiring approximately 15 s, while eight iterations are completed in 44 s to reach 1.3\% deviation [Fig.~\ref{fig:feedback}(a, b)]. To examine the dependence on the number of addressed NDs, we repeat the optimization progress from uniform-weights for groups of 25, 50, 75, 100, 125, and 139 NDs, selected in descending order of their fluorescence intensity under wide-field illumination. Figure~\ref{fig:feedback}(d) shows that all groups reach an root-mean-square deviation below 5\% within one to three iterations, with little dependence on the number of addressed NDs. The rapid convergence suggests that the feedback approach could be extended to larger ND numbers without substantially increasing the optimization time.
Further Scaling will depend on the available excitation power, which determines the fluorescence signal achievable per ND, and on the density of NDs, which sets how many of them lie within the addressable field of view and whether they remain far enough apart to be addressed individually.

\begin{figure}[t!]
  \centering
  \includegraphics[width=\columnwidth]{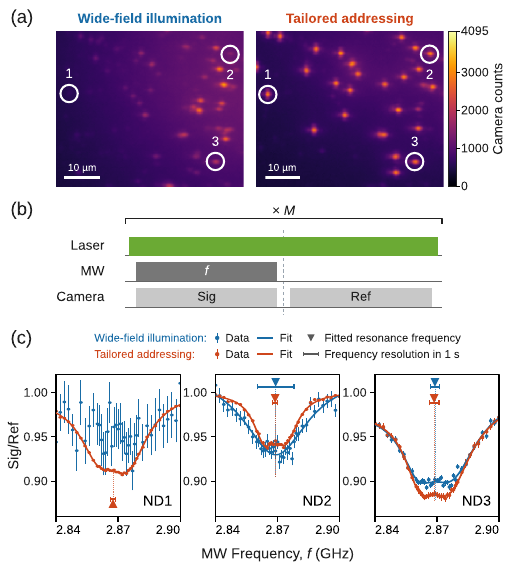}
\caption{\textbf{Electron spin resonance of individual nanodiamonds under wide-field illumination and tailored addressing.}
(a) Fluorescence images of the same region under the two illumination conditions, displayed on the same color scale.
Circles mark the three representative nanodiamonds (ND1-ND3) analyzed in (c).
(b) Measurement sequence. Under continuous laser excitation, the camera alternately records signal (Sig) and reference (Ref) frames with the microwave (MW) on and off, respectively. Each trial is repeated $M$ times at each MW frequency.
(c) Electron spin spectra of ND1-ND3 under wide-field illumination (blue) and tailored addressing (red).
Points show the ratio of the mean signal to mean reference fluorescence (Sig/Ref), with error bars indicating the standard error over $M=200$ measurements. Solid lines are Lorentzian fits.
Triangles mark the fitted resonance frequencies, and the half-widths of the horizontal bars indicate the corresponding estimated frequency resolution for 1 s of camera exposure.
}
  \label{fig:esr}
\end{figure}

\begin{figure*}[t!]
  \centering
  \includegraphics[width=\textwidth]{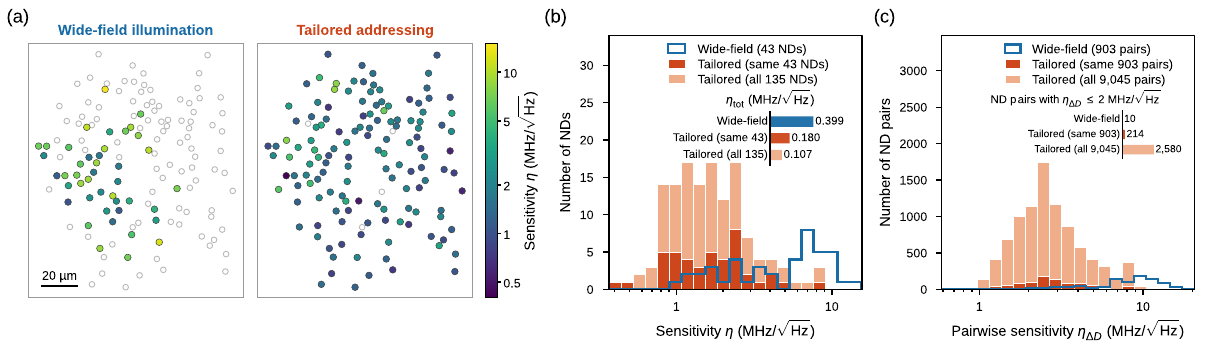}
\caption{\textbf{Enhanced parallel quantum sensing with intensity-tailored addressing.}
(a) Spatial maps of the sensitivity $\eta$ under wide-field illumination and tailored addressing. Filled circles indicate NDs with resolvable ESR spectra (43 and 135 out of 139, respectively), colored by their sensitivity on a logarithmic scale. Open circles indicate the NDs without resolvable ESR spectra.
(b) Distributions of $\eta$ for the 43 NDs under wide-field illumination (blue), the same 43 NDs under tailored addressing (red), and all 135 NDs under tailored addressing (orange). Inset: total sensitivity  $\eta_{\mathrm{tot}}$ to a resonance-frequency shift common to the NDs in each group.
(c) Distributions of the pairwise sensitivity $\eta_{\Delta D}$ to resonance-frequency differences for the corresponding groups of 903, 903, and 9045 ND pairs. Inset: number of ND pairs with $\eta_{\Delta D}\leq 2~\mathrm{MHz}/\sqrt{\mathrm{Hz}}$. The total and pairwise sensitivities are calculated assuming independent measurement noise between NDs.}
  \label{fig:sens}
\end{figure*}

\textit{Enhanced Parallel Quantum Sensing}---To evaluate whether tailored addressing improves the sensing performance, we compare electron spin resonance (ESR) measurements of the same 139 NDs under wide-field illumination and under tailored addressing (Fig.~\ref{fig:esr}).
The measurements are performed at zero applied magnetic field under continuous 532-nm excitation. At each microwave frequencies $f$, the camera alternately records signal and reference frames with the microwave on and off, respectively, repeating 200 times [Fig.~\ref{fig:esr}(b)]. The normalized ESR response is obtained from the ratio of the mean signal to reference fluorescence. The two illumination conditions use exposure times of 32.59 and 43.00 ms, respectively, with identical microwave settings and data analysis.

Figure~\ref{fig:esr}(c) compares the ESR spectra of three representative NDs under the two illumination conditions. For ND1, no clear resonance is observed under wide-field illumination, whereas tailored addressing reveals a distinct ESR signal. ND2 exhibits a more clearly resolved resonance under tailored addressing, while ND3 shows little change between the two conditions. To quantify the sensing performance, we fit the ESR spectra with one or two Lorentzian dips and evaluate the sensitivity to a shift in resonance frequency,
\begin{equation}
    \eta=\frac{\sigma\sqrt{2\tau}}{|\frac{\partial L}{\partial f}|_{\mathrm{max}}},
\end{equation}
where $\sigma$ is the single-pair noise of the normalized fluorescence signal (Sig/Ref), $\tau$ is the exposure time per frame, and $L(f)$ is the fitted ESR response. The factor $2\tau$ is the exposure time of one such pair, which consists of one signal frame and one reference frame.
The sensitivity is normalized to 1 s of camera exposure, excluding camera dead time (see Supplementary Information for fitting and analysis details). Under tailored addressing, ND1 achieves a sensitivity of $1.2~\mathrm{MHz}/\sqrt{\mathrm{Hz}}$. For ND2, the sensitivity improves from 8.8 to 1.2 $\mathrm{MHz}/\sqrt{\mathrm{Hz}}$, a factor of more than seven, while ND3 maintains a comparable sensitivity of 2.3 $\mathrm{MHz}/\sqrt{\mathrm{Hz}}$ under both conditions. The complete ESR spectra and corresponding fitting results for all 139 NDs are provided in the Supplementary Information.

We next evaluate the sensing performance across all 139 NDs (Fig.~\ref{fig:sens}). We identify NDs with resolvable ESR spectra using the same criteria for resonance contrast, fitting quality, frequency uncertainty, and linewidth under both illumination conditions (see Supplementary Information). The number of such NDs increases from 43 under wide-field illumination to 135 under tailored addressing [Fig.~\ref{fig:sens}(a)]. All 43 NDs remain resolvable, while 92 additional NDs become accessible. For the same 43 NDs, the median sensitivity improves from 6.0 to 1.7 $\mathrm{MHz}/\sqrt{\mathrm{Hz}}$, with a median individual improvement factor of 3.1 [Fig.~\ref{fig:sens}(b)]. The 92 newly accessible NDs exhibit a median sensitivity of 1.5 $\mathrm{MHz}/\sqrt{\mathrm{Hz}}$, comparable to that of the original 43 NDs under tailored addressing. Thus, tailored addressing improves the sensitivity of individual sensors while substantially increasing the number available for parallel measurements.

We further evaluate the total sensitivity to a resonance-frequency shift common to all NDs. Assuming independent measurement noise, the total sensitivity is $\eta_{\mathrm{tot}}=(\sum_i\eta_i^{-2})^{-1/2}$, where the sum runs over NDs with resolvable ESR spectra. Under wide-field illumination, the 43 NDs yield a total sensitivity of 0.399 $\mathrm{MHz}/\sqrt{\mathrm{Hz}}$. Tailored addressing improves this value to 0.180 $\mathrm{MHz}/\sqrt{\mathrm{Hz}}$ for the same 43 NDs and to 0.107 $\mathrm{MHz}/\sqrt{\mathrm{Hz}}$ when all 135 NDs are included [inset of Fig.~\ref{fig:sens}(b)]. The latter corresponds to a 3.7-fold improvement over wide-field illumination.

Beyond sensing a common frequency shift, parallel readout can reveal spatial variations through comparisons between NDs. To quantify how precisely such variations can be resolved, we characterize each ND pair $(i,j)$ by its sensitivity to the resonance-frequency difference, $\eta_{\Delta D}^{(ij)}=(\eta_i^2+\eta_j^2)^{1/2}$, assuming independent measurement noise. Unlike the total sensitivity to a common shift, which is dominated by the most sensitive NDs, the pairwise sensitivity is limited by the less sensitive member of each pair.
Under wide-field illumination, the 43 resolvable NDs provide 903 pairs with a median pairwise sensitivity of 8.8 $\mathrm{MHz}/\sqrt{\mathrm{Hz}}$ [Fig.~\ref{fig:sens}(c)]. For the same pairs, tailored addressing improves the median sensitivity to 2.6 $\mathrm{MHz}/\sqrt{\mathrm{Hz}}$, a factor of 3.4. Including all 135 NDs increases the number of available pairs tenfold to 9045, with a median sensitivity of 2.5 $\mathrm{MHz}/\sqrt{\mathrm{Hz}}$, a factor of 3.5 better than under wide-field illumination.
Notably, the number of pairs with a sensitivity of 2 $\mathrm{MHz}/\sqrt{\mathrm{Hz}}$ or better increases from 10 under wide-field illumination to 2580 under tailored addressing. These results demonstrate that tailored addressing enhances both the precision and spatial coverage of parallel quantum sensing.

\textit{Discussion and Outlook}---We have demonstrated enhanced parallel quantum sensing through intensity-tailored optical addressing of NV centers in spatially distributed nanodiamonds. By combining holographic excitation with fluorescence feedback, we independently adjust the excitation intensity of each ND to compensate for variations in its optical response. For 139 NDs, tailored addressing increases the number of sensors with resolvable ESR spectra from 43 to 135, despite the diffraction losses associated with holographic excitation. The total sensitivity  to a common resonance-frequency shift improves by a factor of 3.7, while the number of available ND pairs for spatial comparisons increases tenfold. These results demonstrate that tailored optical excitation can substantially improve the utilization of individual sensors and the performance of parallel quantum sensing.

The approach offers several directions for further development. Increasing the number of simultaneously addressed NDs will require overcoming limitations imposed by the available optical power, hologram diffraction efficiency, and spatial separation between sensors~\cite{Ronzitti2012slm,CatalaCastro2021positioning}. Faster hologram generation and feedback could further improve the optimization speed.
The feedback could also use quantities other than fluorescence. With the ESR contrast, linewidth, or sensitivity of each ND as the feedback signal, the excitation intensities would be optimized directly for sensing performance~\cite{Barry2020sensitivity}. Such optimization could target individual sensor sensitivities or prioritize selected ND pairs for spatially differential measurements.

Intensity-tailored addressing also offers opportunities for biological sensing.
Concentrating excitation light at the ND locations may reduce unnecessary illumination of the surrounding biological sample~\cite{Icha2017phototoxicity,Laissue2017phototoxicity}.
For temperature differences between positions inside a single living cell~\cite{Kucsko2013thermometry}, tailored addressing would allow more positions to be compared and each comparison to be made with higher precision.
For a temperature change shared by a whole organism, such as its thermogenic response to a chemical stimulus~\cite{Fujiwara2020real}, more resolvable NDs would contribute to the total sensitivity.
In addition, the present feedback speed could accommodate slowly varying ND configurations, while faster updates and adaptive spot tracking would be needed for more dynamic environments ~\cite{McGuinness2011tracking,Fujiwara2020real}. These developments could enable multiplexed thermometry and spatially resolved quantum sensing using NDs in living cells and other complex biological systems~\cite{Kucsko2013thermometry,Nishimura2021wide,Lee2025Organelle}.

\par\vspace{0.5\baselineskip}
\vspace{6pt}

\textit{Acknowledgments---} We thank Jun-ji Chen and Hua-yu Yang for helpful discussions. We acknowledge the support from the Quantum Science and Technology National Science and Technology Major Project (grant nos. 2021ZD0302203, 2021ZD0301601, 2021ZD0301605 and 2024ZD0301800), the National Natural Science
Foundation of China (grant nos. T2225008, 12474214 and 123B2072), the Tsinghua University Dushi Program, the Shanghai Qi Zhi Institute (grant nos. SQZ202317, and SQZ202318), the Fundamental and Interdisciplinary Disciplines Breakthrough Plan of the Ministry of Education of China (JYB2025XDXM112), and the Tsinghua
University Initiative Scientific Research Program. P.-Y. H. acknowledges the start-up fund from Tsinghua University.

\par\vspace{0.5\baselineskip}
\vspace{6pt}

\par\vspace{0.5\baselineskip}
\vspace{6pt}

\textit{Data availability---} The data produced in this work are available from the corresponding authors upon reasonable request.

\bibliographystyle{apsrev4-2}
\bibliography{references}

\begin{thebibliography}{46}%
\makeatletter
\providecommand \@ifxundefined [1]{%
 \@ifx{#1\undefined}
}%
\providecommand \@ifnum [1]{%
 \ifnum #1\expandafter \@firstoftwo
 \else \expandafter \@secondoftwo
 \fi
}%
\providecommand \@ifx [1]{%
 \ifx #1\expandafter \@firstoftwo
 \else \expandafter \@secondoftwo
 \fi
}%
\providecommand \natexlab [1]{#1}%
\providecommand \enquote  [1]{``#1''}%
\providecommand \bibnamefont  [1]{#1}%
\providecommand \bibfnamefont [1]{#1}%
\providecommand \citenamefont [1]{#1}%
\providecommand \href@noop [0]{\@secondoftwo}%
\providecommand \href [0]{\begingroup \@sanitize@url \@href}%
\providecommand \@href[1]{\@@startlink{#1}\@@href}%
\providecommand \@@href[1]{\endgroup#1\@@endlink}%
\providecommand \@sanitize@url [0]{\catcode `\\12\catcode `\$12\catcode `\&12\catcode `\#12\catcode `\^12\catcode `\_12\catcode `\%12\relax}%
\providecommand \@@startlink[1]{}%
\providecommand \@@endlink[0]{}%
\providecommand \url  [0]{\begingroup\@sanitize@url \@url }%
\providecommand \@url [1]{\endgroup\@href {#1}{\urlprefix }}%
\providecommand \urlprefix  [0]{URL }%
\providecommand \Eprint [0]{\href }%
\providecommand \doibase [0]{https://doi.org/}%
\providecommand \selectlanguage [0]{\@gobble}%
\providecommand \bibinfo  [0]{\@secondoftwo}%
\providecommand \bibfield  [0]{\@secondoftwo}%
\providecommand \translation [1]{[#1]}%
\providecommand \BibitemOpen [0]{}%
\providecommand \bibitemStop [0]{}%
\providecommand \bibitemNoStop [0]{.\EOS\space}%
\providecommand \EOS [0]{\spacefactor3000\relax}%
\providecommand \BibitemShut  [1]{\csname bibitem#1\endcsname}%
\let\auto@bib@innerbib\@empty
%</preamble>
\bibitem [{\citenamefont {Taylor}\ \emph {et~al.}(2008)\citenamefont {Taylor}, \citenamefont {Cappellaro}, \citenamefont {Childress}, \citenamefont {Jiang}, \citenamefont {Budker}, \citenamefont {Hemmer}, \citenamefont {Yacoby}, \citenamefont {Walsworth},\ and\ \citenamefont {Lukin}}]{Taylor2008magnetometer}%
  \BibitemOpen
  \bibfield  {author} {\bibinfo {author} {\bibfnamefont {J.~M.}\ \bibnamefont {Taylor}}, \bibinfo {author} {\bibfnamefont {P.}~\bibnamefont {Cappellaro}}, \bibinfo {author} {\bibfnamefont {L.}~\bibnamefont {Childress}}, \bibinfo {author} {\bibfnamefont {L.}~\bibnamefont {Jiang}}, \bibinfo {author} {\bibfnamefont {D.}~\bibnamefont {Budker}}, \bibinfo {author} {\bibfnamefont {P.~R.}\ \bibnamefont {Hemmer}}, \bibinfo {author} {\bibfnamefont {A.}~\bibnamefont {Yacoby}}, \bibinfo {author} {\bibfnamefont {R.}~\bibnamefont {Walsworth}},\ and\ \bibinfo {author} {\bibfnamefont {M.~D.}\ \bibnamefont {Lukin}},\ }\href {https://doi.org/10.1038/nphys1075} {\bibfield  {journal} {\bibinfo  {journal} {Nature Physics}\ }\textbf {\bibinfo {volume} {4}},\ \bibinfo {pages} {810} (\bibinfo {year} {2008})}\BibitemShut {NoStop}%
\bibitem [{\citenamefont {Maze}\ \emph {et~al.}(2008)\citenamefont {Maze}, \citenamefont {Stanwix}, \citenamefont {Hodges}, \citenamefont {Hong}, \citenamefont {Taylor}, \citenamefont {Cappellaro}, \citenamefont {Jiang}, \citenamefont {Gurudev~Dutt}, \citenamefont {Togan}, \citenamefont {Zibrov}, \citenamefont {Yacoby}, \citenamefont {Walsworth},\ and\ \citenamefont {Lukin}}]{Maze2008magnetic}%
  \BibitemOpen
  \bibfield  {author} {\bibinfo {author} {\bibfnamefont {J.~R.}\ \bibnamefont {Maze}}, \bibinfo {author} {\bibfnamefont {P.~L.}\ \bibnamefont {Stanwix}}, \bibinfo {author} {\bibfnamefont {J.~S.}\ \bibnamefont {Hodges}}, \bibinfo {author} {\bibfnamefont {S.}~\bibnamefont {Hong}}, \bibinfo {author} {\bibfnamefont {J.~M.}\ \bibnamefont {Taylor}}, \bibinfo {author} {\bibfnamefont {P.}~\bibnamefont {Cappellaro}}, \bibinfo {author} {\bibfnamefont {L.}~\bibnamefont {Jiang}}, \bibinfo {author} {\bibfnamefont {M.~V.}\ \bibnamefont {Gurudev~Dutt}}, \bibinfo {author} {\bibfnamefont {E.}~\bibnamefont {Togan}}, \bibinfo {author} {\bibfnamefont {A.~S.}\ \bibnamefont {Zibrov}}, \bibinfo {author} {\bibfnamefont {A.}~\bibnamefont {Yacoby}}, \bibinfo {author} {\bibfnamefont {R.~L.}\ \bibnamefont {Walsworth}},\ and\ \bibinfo {author} {\bibfnamefont {M.~D.}\ \bibnamefont {Lukin}},\ }\href {https://doi.org/10.1038/nature07279} {\bibfield  {journal} {\bibinfo  {journal} {Nature}\ }\textbf {\bibinfo {volume} {455}},\ \bibinfo
  {pages} {644} (\bibinfo {year} {2008})}\BibitemShut {NoStop}%
\bibitem [{\citenamefont {Balasubramanian}\ \emph {et~al.}(2008)\citenamefont {Balasubramanian}, \citenamefont {Chan}, \citenamefont {Kolesov}, \citenamefont {Al-Hmoud}, \citenamefont {Tisler}, \citenamefont {Shin}, \citenamefont {Kim}, \citenamefont {Wojcik}, \citenamefont {Hemmer}, \citenamefont {Krueger}, \citenamefont {Hanke}, \citenamefont {Leitenstorfer}, \citenamefont {Bratschitsch}, \citenamefont {Jelezko},\ and\ \citenamefont {Wrachtrup}}]{Balasubramanian2008imaging}%
  \BibitemOpen
  \bibfield  {author} {\bibinfo {author} {\bibfnamefont {G.}~\bibnamefont {Balasubramanian}}, \bibinfo {author} {\bibfnamefont {I.~Y.}\ \bibnamefont {Chan}}, \bibinfo {author} {\bibfnamefont {R.}~\bibnamefont {Kolesov}}, \bibinfo {author} {\bibfnamefont {M.}~\bibnamefont {Al-Hmoud}}, \bibinfo {author} {\bibfnamefont {J.}~\bibnamefont {Tisler}}, \bibinfo {author} {\bibfnamefont {C.}~\bibnamefont {Shin}}, \bibinfo {author} {\bibfnamefont {C.}~\bibnamefont {Kim}}, \bibinfo {author} {\bibfnamefont {A.}~\bibnamefont {Wojcik}}, \bibinfo {author} {\bibfnamefont {P.~R.}\ \bibnamefont {Hemmer}}, \bibinfo {author} {\bibfnamefont {A.}~\bibnamefont {Krueger}}, \bibinfo {author} {\bibfnamefont {T.}~\bibnamefont {Hanke}}, \bibinfo {author} {\bibfnamefont {A.}~\bibnamefont {Leitenstorfer}}, \bibinfo {author} {\bibfnamefont {R.}~\bibnamefont {Bratschitsch}}, \bibinfo {author} {\bibfnamefont {F.}~\bibnamefont {Jelezko}},\ and\ \bibinfo {author} {\bibfnamefont {J.}~\bibnamefont {Wrachtrup}},\ }\href
  {https://doi.org/10.1038/nature07278} {\bibfield  {journal} {\bibinfo  {journal} {Nature}\ }\textbf {\bibinfo {volume} {455}},\ \bibinfo {pages} {648} (\bibinfo {year} {2008})}\BibitemShut {NoStop}%
\bibitem [{\citenamefont {Maletinsky}\ \emph {et~al.}(2012)\citenamefont {Maletinsky}, \citenamefont {Hong}, \citenamefont {Grinolds}, \citenamefont {Hausmann}, \citenamefont {Lukin}, \citenamefont {Walsworth}, \citenamefont {Loncar},\ and\ \citenamefont {Yacoby}}]{Maletinsky2012scanning}%
  \BibitemOpen
  \bibfield  {author} {\bibinfo {author} {\bibfnamefont {P.}~\bibnamefont {Maletinsky}}, \bibinfo {author} {\bibfnamefont {S.}~\bibnamefont {Hong}}, \bibinfo {author} {\bibfnamefont {M.~S.}\ \bibnamefont {Grinolds}}, \bibinfo {author} {\bibfnamefont {B.}~\bibnamefont {Hausmann}}, \bibinfo {author} {\bibfnamefont {M.~D.}\ \bibnamefont {Lukin}}, \bibinfo {author} {\bibfnamefont {R.~L.}\ \bibnamefont {Walsworth}}, \bibinfo {author} {\bibfnamefont {M.}~\bibnamefont {Loncar}},\ and\ \bibinfo {author} {\bibfnamefont {A.}~\bibnamefont {Yacoby}},\ }\href {https://doi.org/10.1038/nnano.2012.50} {\bibfield  {journal} {\bibinfo  {journal} {Nature Nanotechnology}\ }\textbf {\bibinfo {volume} {7}},\ \bibinfo {pages} {320} (\bibinfo {year} {2012})}\BibitemShut {NoStop}%
\bibitem [{\citenamefont {Grinolds}\ \emph {et~al.}(2013)\citenamefont {Grinolds}, \citenamefont {Hong}, \citenamefont {Maletinsky}, \citenamefont {Luan}, \citenamefont {Lukin}, \citenamefont {Walsworth},\ and\ \citenamefont {Yacoby}}]{Grinolds2013singleSpin}%
  \BibitemOpen
  \bibfield  {author} {\bibinfo {author} {\bibfnamefont {M.~S.}\ \bibnamefont {Grinolds}}, \bibinfo {author} {\bibfnamefont {S.}~\bibnamefont {Hong}}, \bibinfo {author} {\bibfnamefont {P.}~\bibnamefont {Maletinsky}}, \bibinfo {author} {\bibfnamefont {L.}~\bibnamefont {Luan}}, \bibinfo {author} {\bibfnamefont {M.~D.}\ \bibnamefont {Lukin}}, \bibinfo {author} {\bibfnamefont {R.~L.}\ \bibnamefont {Walsworth}},\ and\ \bibinfo {author} {\bibfnamefont {A.}~\bibnamefont {Yacoby}},\ }\href {https://doi.org/10.1038/nphys2543} {\bibfield  {journal} {\bibinfo  {journal} {Nature Physics}\ }\textbf {\bibinfo {volume} {9}},\ \bibinfo {pages} {215} (\bibinfo {year} {2013})}\BibitemShut {NoStop}%
\bibitem [{\citenamefont {Acosta}\ \emph {et~al.}(2010)\citenamefont {Acosta}, \citenamefont {Bauch}, \citenamefont {Ledbetter}, \citenamefont {Waxman}, \citenamefont {Bouchard},\ and\ \citenamefont {Budker}}]{Acosta2010temperature}%
  \BibitemOpen
  \bibfield  {author} {\bibinfo {author} {\bibfnamefont {V.~M.}\ \bibnamefont {Acosta}}, \bibinfo {author} {\bibfnamefont {E.}~\bibnamefont {Bauch}}, \bibinfo {author} {\bibfnamefont {M.~P.}\ \bibnamefont {Ledbetter}}, \bibinfo {author} {\bibfnamefont {A.}~\bibnamefont {Waxman}}, \bibinfo {author} {\bibfnamefont {L.-S.}\ \bibnamefont {Bouchard}},\ and\ \bibinfo {author} {\bibfnamefont {D.}~\bibnamefont {Budker}},\ }\href {https://doi.org/10.1103/PhysRevLett.104.070801} {\bibfield  {journal} {\bibinfo  {journal} {Physical Review Letters}\ }\textbf {\bibinfo {volume} {104}},\ \bibinfo {pages} {070801} (\bibinfo {year} {2010})}\BibitemShut {NoStop}%
\bibitem [{\citenamefont {Neumann}\ \emph {et~al.}(2013)\citenamefont {Neumann}, \citenamefont {Jakobi}, \citenamefont {Dolde}, \citenamefont {Burk}, \citenamefont {Reuter}, \citenamefont {Waldherr}, \citenamefont {Honert}, \citenamefont {Wolf}, \citenamefont {Brunner},\ and\ \citenamefont {Shim}}]{Neumann2013thermometry}%
  \BibitemOpen
  \bibfield  {author} {\bibinfo {author} {\bibfnamefont {P.}~\bibnamefont {Neumann}}, \bibinfo {author} {\bibfnamefont {I.}~\bibnamefont {Jakobi}}, \bibinfo {author} {\bibfnamefont {F.}~\bibnamefont {Dolde}}, \bibinfo {author} {\bibfnamefont {C.}~\bibnamefont {Burk}}, \bibinfo {author} {\bibfnamefont {R.}~\bibnamefont {Reuter}}, \bibinfo {author} {\bibfnamefont {G.}~\bibnamefont {Waldherr}}, \bibinfo {author} {\bibfnamefont {J.}~\bibnamefont {Honert}}, \bibinfo {author} {\bibfnamefont {T.}~\bibnamefont {Wolf}}, \bibinfo {author} {\bibfnamefont {A.}~\bibnamefont {Brunner}},\ and\ \bibinfo {author} {\bibfnamefont {J.~H.}\ \bibnamefont {Shim}},\ }\href {https://doi.org/10.1021/nl401216y} {\bibfield  {journal} {\bibinfo  {journal} {Nano Letters}\ }\textbf {\bibinfo {volume} {13}},\ \bibinfo {pages} {2738} (\bibinfo {year} {2013})}\BibitemShut {NoStop}%
\bibitem [{\citenamefont {Toyli}\ \emph {et~al.}(2013)\citenamefont {Toyli}, \citenamefont {de~las Casas}, \citenamefont {Christle}, \citenamefont {Dobrovitski},\ and\ \citenamefont {Awschalom}}]{Toyli2013thermometry}%
  \BibitemOpen
  \bibfield  {author} {\bibinfo {author} {\bibfnamefont {D.~M.}\ \bibnamefont {Toyli}}, \bibinfo {author} {\bibfnamefont {C.~F.}\ \bibnamefont {de~las Casas}}, \bibinfo {author} {\bibfnamefont {D.~J.}\ \bibnamefont {Christle}}, \bibinfo {author} {\bibfnamefont {V.~V.}\ \bibnamefont {Dobrovitski}},\ and\ \bibinfo {author} {\bibfnamefont {D.~D.}\ \bibnamefont {Awschalom}},\ }\href {https://doi.org/10.1073/pnas.1306825110} {\bibfield  {journal} {\bibinfo  {journal} {Proceedings of the National Academy of Sciences}\ }\textbf {\bibinfo {volume} {110}},\ \bibinfo {pages} {8417} (\bibinfo {year} {2013})}\BibitemShut {NoStop}%
\bibitem [{\citenamefont {Kucsko}\ \emph {et~al.}(2013)\citenamefont {Kucsko}, \citenamefont {Maurer}, \citenamefont {Yao}, \citenamefont {Kubo}, \citenamefont {Noh}, \citenamefont {Lo}, \citenamefont {Park},\ and\ \citenamefont {Lukin}}]{Kucsko2013thermometry}%
  \BibitemOpen
  \bibfield  {author} {\bibinfo {author} {\bibfnamefont {G.}~\bibnamefont {Kucsko}}, \bibinfo {author} {\bibfnamefont {P.~C.}\ \bibnamefont {Maurer}}, \bibinfo {author} {\bibfnamefont {N.~Y.}\ \bibnamefont {Yao}}, \bibinfo {author} {\bibfnamefont {M.}~\bibnamefont {Kubo}}, \bibinfo {author} {\bibfnamefont {H.~J.}\ \bibnamefont {Noh}}, \bibinfo {author} {\bibfnamefont {P.~K.}\ \bibnamefont {Lo}}, \bibinfo {author} {\bibfnamefont {H.}~\bibnamefont {Park}},\ and\ \bibinfo {author} {\bibfnamefont {M.~D.}\ \bibnamefont {Lukin}},\ }\href {https://doi.org/10.1038/nature12373} {\bibfield  {journal} {\bibinfo  {journal} {Nature}\ }\textbf {\bibinfo {volume} {500}},\ \bibinfo {pages} {54} (\bibinfo {year} {2013})}\BibitemShut {NoStop}%
\bibitem [{\citenamefont {Barry}\ \emph {et~al.}(2020)\citenamefont {Barry}, \citenamefont {Schloss}, \citenamefont {Bauch}, \citenamefont {Turner}, \citenamefont {Hart}, \citenamefont {Pham},\ and\ \citenamefont {Walsworth}}]{Barry2020sensitivity}%
  \BibitemOpen
  \bibfield  {author} {\bibinfo {author} {\bibfnamefont {J.~F.}\ \bibnamefont {Barry}}, \bibinfo {author} {\bibfnamefont {J.~M.}\ \bibnamefont {Schloss}}, \bibinfo {author} {\bibfnamefont {E.}~\bibnamefont {Bauch}}, \bibinfo {author} {\bibfnamefont {M.~J.}\ \bibnamefont {Turner}}, \bibinfo {author} {\bibfnamefont {C.~A.}\ \bibnamefont {Hart}}, \bibinfo {author} {\bibfnamefont {L.~M.}\ \bibnamefont {Pham}},\ and\ \bibinfo {author} {\bibfnamefont {R.~L.}\ \bibnamefont {Walsworth}},\ }\href {https://doi.org/10.1103/RevModPhys.92.015004} {\bibfield  {journal} {\bibinfo  {journal} {Reviews of Modern Physics}\ }\textbf {\bibinfo {volume} {92}},\ \bibinfo {pages} {015004} (\bibinfo {year} {2020})}\BibitemShut {NoStop}%
\bibitem [{\citenamefont {Le~Sage}\ \emph {et~al.}(2013)\citenamefont {Le~Sage}, \citenamefont {Arai}, \citenamefont {Glenn}, \citenamefont {DeVience}, \citenamefont {Pham}, \citenamefont {Rahn-Lee}, \citenamefont {Lukin}, \citenamefont {Yacoby}, \citenamefont {Komeili},\ and\ \citenamefont {Walsworth}}]{LeSage2013Optical}%
  \BibitemOpen
  \bibfield  {author} {\bibinfo {author} {\bibfnamefont {D.}~\bibnamefont {Le~Sage}}, \bibinfo {author} {\bibfnamefont {K.}~\bibnamefont {Arai}}, \bibinfo {author} {\bibfnamefont {D.~R.}\ \bibnamefont {Glenn}}, \bibinfo {author} {\bibfnamefont {S.~J.}\ \bibnamefont {DeVience}}, \bibinfo {author} {\bibfnamefont {L.~M.}\ \bibnamefont {Pham}}, \bibinfo {author} {\bibfnamefont {L.}~\bibnamefont {Rahn-Lee}}, \bibinfo {author} {\bibfnamefont {M.~D.}\ \bibnamefont {Lukin}}, \bibinfo {author} {\bibfnamefont {A.}~\bibnamefont {Yacoby}}, \bibinfo {author} {\bibfnamefont {A.}~\bibnamefont {Komeili}},\ and\ \bibinfo {author} {\bibfnamefont {R.~L.}\ \bibnamefont {Walsworth}},\ }\href {https://doi.org/10.1038/nature12072} {\bibfield  {journal} {\bibinfo  {journal} {Nature}\ }\textbf {\bibinfo {volume} {496}},\ \bibinfo {pages} {486} (\bibinfo {year} {2013})}\BibitemShut {NoStop}%
\bibitem [{\citenamefont {McGuinness}\ \emph {et~al.}(2011)\citenamefont {McGuinness}, \citenamefont {Yan}, \citenamefont {Stacey}, \citenamefont {Simpson}, \citenamefont {Hall}, \citenamefont {Maclaurin}, \citenamefont {Prawer}, \citenamefont {Mulvaney}, \citenamefont {Wrachtrup}, \citenamefont {Caruso}, \citenamefont {Scholten},\ and\ \citenamefont {Hollenberg}}]{McGuinness2011tracking}%
  \BibitemOpen
  \bibfield  {author} {\bibinfo {author} {\bibfnamefont {L.~P.}\ \bibnamefont {McGuinness}}, \bibinfo {author} {\bibfnamefont {Y.}~\bibnamefont {Yan}}, \bibinfo {author} {\bibfnamefont {A.}~\bibnamefont {Stacey}}, \bibinfo {author} {\bibfnamefont {D.~A.}\ \bibnamefont {Simpson}}, \bibinfo {author} {\bibfnamefont {L.~T.}\ \bibnamefont {Hall}}, \bibinfo {author} {\bibfnamefont {D.}~\bibnamefont {Maclaurin}}, \bibinfo {author} {\bibfnamefont {S.}~\bibnamefont {Prawer}}, \bibinfo {author} {\bibfnamefont {P.}~\bibnamefont {Mulvaney}}, \bibinfo {author} {\bibfnamefont {J.}~\bibnamefont {Wrachtrup}}, \bibinfo {author} {\bibfnamefont {F.}~\bibnamefont {Caruso}}, \bibinfo {author} {\bibfnamefont {R.~E.}\ \bibnamefont {Scholten}},\ and\ \bibinfo {author} {\bibfnamefont {L.~C.~L.}\ \bibnamefont {Hollenberg}},\ }\href {https://doi.org/10.1038/nnano.2011.64} {\bibfield  {journal} {\bibinfo  {journal} {Nature Nanotechnology}\ }\textbf {\bibinfo {volume} {6}},\ \bibinfo {pages} {358} (\bibinfo {year} {2011})}\BibitemShut
  {NoStop}%
\bibitem [{\citenamefont {Fujiwara}\ \emph {et~al.}(2020)\citenamefont {Fujiwara}, \citenamefont {Sun}, \citenamefont {Dohms}, \citenamefont {Nishimura}, \citenamefont {Suto}, \citenamefont {Takezawa}, \citenamefont {Oshimi}, \citenamefont {Zhao}, \citenamefont {Sadzak}, \citenamefont {Umehara}, \citenamefont {Teki}, \citenamefont {Komatsu}, \citenamefont {Benson}, \citenamefont {Shikano},\ and\ \citenamefont {Kage-Nakadai}}]{Fujiwara2020real}%
  \BibitemOpen
  \bibfield  {author} {\bibinfo {author} {\bibfnamefont {M.}~\bibnamefont {Fujiwara}}, \bibinfo {author} {\bibfnamefont {S.}~\bibnamefont {Sun}}, \bibinfo {author} {\bibfnamefont {A.}~\bibnamefont {Dohms}}, \bibinfo {author} {\bibfnamefont {Y.}~\bibnamefont {Nishimura}}, \bibinfo {author} {\bibfnamefont {K.}~\bibnamefont {Suto}}, \bibinfo {author} {\bibfnamefont {Y.}~\bibnamefont {Takezawa}}, \bibinfo {author} {\bibfnamefont {K.}~\bibnamefont {Oshimi}}, \bibinfo {author} {\bibfnamefont {L.}~\bibnamefont {Zhao}}, \bibinfo {author} {\bibfnamefont {N.}~\bibnamefont {Sadzak}}, \bibinfo {author} {\bibfnamefont {Y.}~\bibnamefont {Umehara}}, \bibinfo {author} {\bibfnamefont {Y.}~\bibnamefont {Teki}}, \bibinfo {author} {\bibfnamefont {N.}~\bibnamefont {Komatsu}}, \bibinfo {author} {\bibfnamefont {O.}~\bibnamefont {Benson}}, \bibinfo {author} {\bibfnamefont {Y.}~\bibnamefont {Shikano}},\ and\ \bibinfo {author} {\bibfnamefont {E.}~\bibnamefont {Kage-Nakadai}},\ }\href {https://doi.org/10.1126/sciadv.aba9636} {\bibfield
   {journal} {\bibinfo  {journal} {Science Advances}\ }\textbf {\bibinfo {volume} {6}},\ \bibinfo {pages} {eaba9636} (\bibinfo {year} {2020})}\BibitemShut {NoStop}%
\bibitem [{\citenamefont {Simpson}\ \emph {et~al.}(2017)\citenamefont {Simpson}, \citenamefont {Morrisroe}, \citenamefont {McCoey}, \citenamefont {Lombard}, \citenamefont {Mendis}, \citenamefont {Treussart}, \citenamefont {Hall}, \citenamefont {Petrou},\ and\ \citenamefont {Hollenberg}}]{Simpson2017neuronal}%
  \BibitemOpen
  \bibfield  {author} {\bibinfo {author} {\bibfnamefont {D.~A.}\ \bibnamefont {Simpson}}, \bibinfo {author} {\bibfnamefont {E.}~\bibnamefont {Morrisroe}}, \bibinfo {author} {\bibfnamefont {J.~M.}\ \bibnamefont {McCoey}}, \bibinfo {author} {\bibfnamefont {A.~H.}\ \bibnamefont {Lombard}}, \bibinfo {author} {\bibfnamefont {D.~C.}\ \bibnamefont {Mendis}}, \bibinfo {author} {\bibfnamefont {F.}~\bibnamefont {Treussart}}, \bibinfo {author} {\bibfnamefont {L.~T.}\ \bibnamefont {Hall}}, \bibinfo {author} {\bibfnamefont {S.}~\bibnamefont {Petrou}},\ and\ \bibinfo {author} {\bibfnamefont {L.~C.~L.}\ \bibnamefont {Hollenberg}},\ }\href {https://doi.org/10.1021/acsnano.7b04850} {\bibfield  {journal} {\bibinfo  {journal} {ACS Nano}\ }\textbf {\bibinfo {volume} {11}},\ \bibinfo {pages} {12077} (\bibinfo {year} {2017})}\BibitemShut {NoStop}%
\bibitem [{\citenamefont {Sotoma}\ \emph {et~al.}(2021)\citenamefont {Sotoma}, \citenamefont {Zhong}, \citenamefont {Kah}, \citenamefont {Yamashita}, \citenamefont {Plakhotnik}, \citenamefont {Harada},\ and\ \citenamefont {Suzuki}}]{Sotoma2021insitu}%
  \BibitemOpen
  \bibfield  {author} {\bibinfo {author} {\bibfnamefont {S.}~\bibnamefont {Sotoma}}, \bibinfo {author} {\bibfnamefont {C.}~\bibnamefont {Zhong}}, \bibinfo {author} {\bibfnamefont {J.~C.~Y.}\ \bibnamefont {Kah}}, \bibinfo {author} {\bibfnamefont {H.}~\bibnamefont {Yamashita}}, \bibinfo {author} {\bibfnamefont {T.}~\bibnamefont {Plakhotnik}}, \bibinfo {author} {\bibfnamefont {Y.}~\bibnamefont {Harada}},\ and\ \bibinfo {author} {\bibfnamefont {M.}~\bibnamefont {Suzuki}},\ }\href {https://doi.org/10.1126/sciadv.abd7888} {\bibfield  {journal} {\bibinfo  {journal} {Science Advances}\ }\textbf {\bibinfo {volume} {7}},\ \bibinfo {pages} {eabd7888} (\bibinfo {year} {2021})}\BibitemShut {NoStop}%
\bibitem [{\citenamefont {Nie}\ \emph {et~al.}(2021)\citenamefont {Nie}, \citenamefont {Nusantara}, \citenamefont {Damle}, \citenamefont {Sharmin}, \citenamefont {Evans}, \citenamefont {Hemelaar}, \citenamefont {van~der Laan}, \citenamefont {Li}, \citenamefont {Perona~Martinez}, \citenamefont {Vedelaar}, \citenamefont {Chipaux},\ and\ \citenamefont {Schirhagl}}]{Nie2021Quantum}%
  \BibitemOpen
  \bibfield  {author} {\bibinfo {author} {\bibfnamefont {L.}~\bibnamefont {Nie}}, \bibinfo {author} {\bibfnamefont {A.~C.}\ \bibnamefont {Nusantara}}, \bibinfo {author} {\bibfnamefont {V.~G.}\ \bibnamefont {Damle}}, \bibinfo {author} {\bibfnamefont {R.}~\bibnamefont {Sharmin}}, \bibinfo {author} {\bibfnamefont {E.~P.~P.}\ \bibnamefont {Evans}}, \bibinfo {author} {\bibfnamefont {S.~R.}\ \bibnamefont {Hemelaar}}, \bibinfo {author} {\bibfnamefont {K.~J.}\ \bibnamefont {van~der Laan}}, \bibinfo {author} {\bibfnamefont {R.}~\bibnamefont {Li}}, \bibinfo {author} {\bibfnamefont {F.~P.}\ \bibnamefont {Perona~Martinez}}, \bibinfo {author} {\bibfnamefont {T.}~\bibnamefont {Vedelaar}}, \bibinfo {author} {\bibfnamefont {M.}~\bibnamefont {Chipaux}},\ and\ \bibinfo {author} {\bibfnamefont {R.}~\bibnamefont {Schirhagl}},\ }\href {https://doi.org/10.1126/sciadv.abf0573} {\bibfield  {journal} {\bibinfo  {journal} {Science Advances}\ }\textbf {\bibinfo {volume} {7}},\ \bibinfo {pages} {eabf0573} (\bibinfo {year}
  {2021})}\BibitemShut {NoStop}%
\bibitem [{\citenamefont {Sigaeva}\ \emph {et~al.}(2022)\citenamefont {Sigaeva}, \citenamefont {Shirzad}, \citenamefont {Perona~Martinez}, \citenamefont {Nusantara}, \citenamefont {Mougios}, \citenamefont {Chipaux},\ and\ \citenamefont {Schirhagl}}]{Sigaeva2022Diamond}%
  \BibitemOpen
  \bibfield  {author} {\bibinfo {author} {\bibfnamefont {A.}~\bibnamefont {Sigaeva}}, \bibinfo {author} {\bibfnamefont {H.}~\bibnamefont {Shirzad}}, \bibinfo {author} {\bibfnamefont {F.}~\bibnamefont {Perona~Martinez}}, \bibinfo {author} {\bibfnamefont {A.~C.}\ \bibnamefont {Nusantara}}, \bibinfo {author} {\bibfnamefont {N.}~\bibnamefont {Mougios}}, \bibinfo {author} {\bibfnamefont {M.}~\bibnamefont {Chipaux}},\ and\ \bibinfo {author} {\bibfnamefont {R.}~\bibnamefont {Schirhagl}},\ }\href {https://doi.org/10.1002/smll.202105750} {\bibfield  {journal} {\bibinfo  {journal} {Small}\ }\textbf {\bibinfo {volume} {18}},\ \bibinfo {pages} {2105750} (\bibinfo {year} {2022})}\BibitemShut {NoStop}%
\bibitem [{\citenamefont {Consiglio}\ \emph {et~al.}(2026)\citenamefont {Consiglio}, \citenamefont {Boltyanskiy}, \citenamefont {Mindarava}, \citenamefont {Laube}, \citenamefont {Knolle}, \citenamefont {Berishaj}, \citenamefont {Jelezko},\ and\ \citenamefont {Keshari}}]{Consiglio2026Optically}%
  \BibitemOpen
  \bibfield  {author} {\bibinfo {author} {\bibfnamefont {J.~C.}\ \bibnamefont {Consiglio}}, \bibinfo {author} {\bibfnamefont {R.}~\bibnamefont {Boltyanskiy}}, \bibinfo {author} {\bibfnamefont {Y.~L.}\ \bibnamefont {Mindarava}}, \bibinfo {author} {\bibfnamefont {C.}~\bibnamefont {Laube}}, \bibinfo {author} {\bibfnamefont {W.}~\bibnamefont {Knolle}}, \bibinfo {author} {\bibfnamefont {M.}~\bibnamefont {Berishaj}}, \bibinfo {author} {\bibfnamefont {F.}~\bibnamefont {Jelezko}},\ and\ \bibinfo {author} {\bibfnamefont {K.~R.}\ \bibnamefont {Keshari}},\ }\href {https://doi.org/10.1021/acsanm.5c05655} {\bibfield  {journal} {\bibinfo  {journal} {ACS Applied Nano Materials}\ }\textbf {\bibinfo {volume} {9}},\ \bibinfo {pages} {3590} (\bibinfo {year} {2026})}\BibitemShut {NoStop}%
\bibitem [{\citenamefont {Miller}\ \emph {et~al.}(2020)\citenamefont {Miller}, \citenamefont {Bezinge}, \citenamefont {Gliddon}, \citenamefont {Huang}, \citenamefont {Dold}, \citenamefont {Gray}, \citenamefont {Heaney}, \citenamefont {Dobson}, \citenamefont {Nastouli}, \citenamefont {Morton},\ and\ \citenamefont {McKendry}}]{Miller2020spin}%
  \BibitemOpen
  \bibfield  {author} {\bibinfo {author} {\bibfnamefont {B.~S.}\ \bibnamefont {Miller}}, \bibinfo {author} {\bibfnamefont {L.}~\bibnamefont {Bezinge}}, \bibinfo {author} {\bibfnamefont {H.~D.}\ \bibnamefont {Gliddon}}, \bibinfo {author} {\bibfnamefont {D.}~\bibnamefont {Huang}}, \bibinfo {author} {\bibfnamefont {G.}~\bibnamefont {Dold}}, \bibinfo {author} {\bibfnamefont {E.~R.}\ \bibnamefont {Gray}}, \bibinfo {author} {\bibfnamefont {J.}~\bibnamefont {Heaney}}, \bibinfo {author} {\bibfnamefont {P.~J.}\ \bibnamefont {Dobson}}, \bibinfo {author} {\bibfnamefont {E.}~\bibnamefont {Nastouli}}, \bibinfo {author} {\bibfnamefont {J.~J.~L.}\ \bibnamefont {Morton}},\ and\ \bibinfo {author} {\bibfnamefont {R.~A.}\ \bibnamefont {McKendry}},\ }\href {https://doi.org/10.1038/s41586-020-2917-1} {\bibfield  {journal} {\bibinfo  {journal} {Nature}\ }\textbf {\bibinfo {volume} {587}},\ \bibinfo {pages} {588} (\bibinfo {year} {2020})}\BibitemShut {NoStop}%
\bibitem [{\citenamefont {Steinert}\ \emph {et~al.}(2013)\citenamefont {Steinert}, \citenamefont {Ziem}, \citenamefont {Hall}, \citenamefont {Zappe}, \citenamefont {Schweikert}, \citenamefont {G{\"o}tz}, \citenamefont {Aird}, \citenamefont {Balasubramanian}, \citenamefont {Hollenberg},\ and\ \citenamefont {Wrachtrup}}]{Steinert2013Magnetic}%
  \BibitemOpen
  \bibfield  {author} {\bibinfo {author} {\bibfnamefont {S.}~\bibnamefont {Steinert}}, \bibinfo {author} {\bibfnamefont {F.}~\bibnamefont {Ziem}}, \bibinfo {author} {\bibfnamefont {L.~T.}\ \bibnamefont {Hall}}, \bibinfo {author} {\bibfnamefont {A.}~\bibnamefont {Zappe}}, \bibinfo {author} {\bibfnamefont {M.}~\bibnamefont {Schweikert}}, \bibinfo {author} {\bibfnamefont {N.}~\bibnamefont {G{\"o}tz}}, \bibinfo {author} {\bibfnamefont {A.}~\bibnamefont {Aird}}, \bibinfo {author} {\bibfnamefont {G.}~\bibnamefont {Balasubramanian}}, \bibinfo {author} {\bibfnamefont {L.}~\bibnamefont {Hollenberg}},\ and\ \bibinfo {author} {\bibfnamefont {J.}~\bibnamefont {Wrachtrup}},\ }\href {https://doi.org/10.1038/ncomms2588} {\bibfield  {journal} {\bibinfo  {journal} {Nature Communications}\ }\textbf {\bibinfo {volume} {4}},\ \bibinfo {pages} {1607} (\bibinfo {year} {2013})}\BibitemShut {NoStop}%
\bibitem [{\citenamefont {Nishimura}\ \emph {et~al.}(2021)\citenamefont {Nishimura}, \citenamefont {Oshimi}, \citenamefont {Umehara}, \citenamefont {Kumon}, \citenamefont {Miyaji}, \citenamefont {Yukawa}, \citenamefont {Shikano}, \citenamefont {Matsubara}, \citenamefont {Fujiwara}, \citenamefont {Baba},\ and\ \citenamefont {Teki}}]{Nishimura2021wide}%
  \BibitemOpen
  \bibfield  {author} {\bibinfo {author} {\bibfnamefont {Y.}~\bibnamefont {Nishimura}}, \bibinfo {author} {\bibfnamefont {K.}~\bibnamefont {Oshimi}}, \bibinfo {author} {\bibfnamefont {Y.}~\bibnamefont {Umehara}}, \bibinfo {author} {\bibfnamefont {Y.}~\bibnamefont {Kumon}}, \bibinfo {author} {\bibfnamefont {K.}~\bibnamefont {Miyaji}}, \bibinfo {author} {\bibfnamefont {H.}~\bibnamefont {Yukawa}}, \bibinfo {author} {\bibfnamefont {Y.}~\bibnamefont {Shikano}}, \bibinfo {author} {\bibfnamefont {T.}~\bibnamefont {Matsubara}}, \bibinfo {author} {\bibfnamefont {M.}~\bibnamefont {Fujiwara}}, \bibinfo {author} {\bibfnamefont {Y.}~\bibnamefont {Baba}},\ and\ \bibinfo {author} {\bibfnamefont {Y.}~\bibnamefont {Teki}},\ }\href {https://doi.org/10.1038/s41598-021-83285-y} {\bibfield  {journal} {\bibinfo  {journal} {Scientific Reports}\ }\textbf {\bibinfo {volume} {11}},\ \bibinfo {pages} {4248} (\bibinfo {year} {2021})}\BibitemShut {NoStop}%
\bibitem [{\citenamefont {Sow}\ \emph {et~al.}(2020)\citenamefont {Sow}, \citenamefont {Steuer}, \citenamefont {Adekanye}, \citenamefont {Gin{\'e}s}, \citenamefont {Mandal}, \citenamefont {Gilboa}, \citenamefont {Williams}, \citenamefont {Smith},\ and\ \citenamefont {Kapanidis}}]{Sow2020high}%
  \BibitemOpen
  \bibfield  {author} {\bibinfo {author} {\bibfnamefont {M.}~\bibnamefont {Sow}}, \bibinfo {author} {\bibfnamefont {H.}~\bibnamefont {Steuer}}, \bibinfo {author} {\bibfnamefont {S.}~\bibnamefont {Adekanye}}, \bibinfo {author} {\bibfnamefont {L.}~\bibnamefont {Gin{\'e}s}}, \bibinfo {author} {\bibfnamefont {S.}~\bibnamefont {Mandal}}, \bibinfo {author} {\bibfnamefont {B.}~\bibnamefont {Gilboa}}, \bibinfo {author} {\bibfnamefont {O.~A.}\ \bibnamefont {Williams}}, \bibinfo {author} {\bibfnamefont {J.~M.}\ \bibnamefont {Smith}},\ and\ \bibinfo {author} {\bibfnamefont {A.~N.}\ \bibnamefont {Kapanidis}},\ }\href {https://doi.org/10.1039/D0NR05931E} {\bibfield  {journal} {\bibinfo  {journal} {Nanoscale}\ }\textbf {\bibinfo {volume} {12}},\ \bibinfo {pages} {21821} (\bibinfo {year} {2020})}\BibitemShut {NoStop}%
\bibitem [{\citenamefont {Lee}\ \emph {et~al.}(2025)\citenamefont {Lee}, \citenamefont {Kim}, \citenamefont {Kim},\ and\ \citenamefont {Lee}}]{Lee2025Organelle}%
  \BibitemOpen
  \bibfield  {author} {\bibinfo {author} {\bibfnamefont {Y.}~\bibnamefont {Lee}}, \bibinfo {author} {\bibfnamefont {K.}~\bibnamefont {Kim}}, \bibinfo {author} {\bibfnamefont {D.}~\bibnamefont {Kim}},\ and\ \bibinfo {author} {\bibfnamefont {J.~S.}\ \bibnamefont {Lee}},\ }\href {https://doi.org/10.1021/jacs.4c16365} {\bibfield  {journal} {\bibinfo  {journal} {Journal of the American Chemical Society}\ }\textbf {\bibinfo {volume} {147}},\ \bibinfo {pages} {13180} (\bibinfo {year} {2025})}\BibitemShut {NoStop}%
\bibitem [{\citenamefont {Xu}\ \emph {et~al.}(2025)\citenamefont {Xu}, \citenamefont {Yang}, \citenamefont {Chen}, \citenamefont {Zhu}, \citenamefont {Lyu}, \citenamefont {Zhang}, \citenamefont {Huang},\ and\ \citenamefont {Feng}}]{Xu2025Multiphysiology}%
  \BibitemOpen
  \bibfield  {author} {\bibinfo {author} {\bibfnamefont {Y.}~\bibnamefont {Xu}}, \bibinfo {author} {\bibfnamefont {Y.}~\bibnamefont {Yang}}, \bibinfo {author} {\bibfnamefont {Y.}~\bibnamefont {Chen}}, \bibinfo {author} {\bibfnamefont {W.}~\bibnamefont {Zhu}}, \bibinfo {author} {\bibfnamefont {S.}~\bibnamefont {Lyu}}, \bibinfo {author} {\bibfnamefont {C.}~\bibnamefont {Zhang}}, \bibinfo {author} {\bibfnamefont {X.}~\bibnamefont {Huang}},\ and\ \bibinfo {author} {\bibfnamefont {J.}~\bibnamefont {Feng}},\ }\href {https://doi.org/10.1093/nsr/nwaf130} {\bibfield  {journal} {\bibinfo  {journal} {National Science Review}\ }\textbf {\bibinfo {volume} {12}},\ \bibinfo {pages} {nwaf130} (\bibinfo {year} {2025})}\BibitemShut {NoStop}%
\bibitem [{\citenamefont {Scholten}\ \emph {et~al.}(2021)\citenamefont {Scholten}, \citenamefont {Healey}, \citenamefont {Robertson}, \citenamefont {Abrahams}, \citenamefont {Broadway},\ and\ \citenamefont {Tetienne}}]{Scholten2021widefield}%
  \BibitemOpen
  \bibfield  {author} {\bibinfo {author} {\bibfnamefont {S.~C.}\ \bibnamefont {Scholten}}, \bibinfo {author} {\bibfnamefont {A.~J.}\ \bibnamefont {Healey}}, \bibinfo {author} {\bibfnamefont {I.~O.}\ \bibnamefont {Robertson}}, \bibinfo {author} {\bibfnamefont {G.~J.}\ \bibnamefont {Abrahams}}, \bibinfo {author} {\bibfnamefont {D.~A.}\ \bibnamefont {Broadway}},\ and\ \bibinfo {author} {\bibfnamefont {J.-P.}\ \bibnamefont {Tetienne}},\ }\href {https://doi.org/10.1063/5.0066733} {\bibfield  {journal} {\bibinfo  {journal} {Journal of Applied Physics}\ }\textbf {\bibinfo {volume} {130}},\ \bibinfo {pages} {150902} (\bibinfo {year} {2021})}\BibitemShut {NoStop}%
\bibitem [{\citenamefont {Igarashi}\ \emph {et~al.}(2012)\citenamefont {Igarashi}, \citenamefont {Yoshinari}, \citenamefont {Yokota}, \citenamefont {Sugi}, \citenamefont {Sugihara}, \citenamefont {Ikeda}, \citenamefont {Sumiya}, \citenamefont {Tsuji}, \citenamefont {Mori}, \citenamefont {Tochio}, \citenamefont {Harada},\ and\ \citenamefont {Shirakawa}}]{Igarashi2012real}%
  \BibitemOpen
  \bibfield  {author} {\bibinfo {author} {\bibfnamefont {R.}~\bibnamefont {Igarashi}}, \bibinfo {author} {\bibfnamefont {Y.}~\bibnamefont {Yoshinari}}, \bibinfo {author} {\bibfnamefont {H.}~\bibnamefont {Yokota}}, \bibinfo {author} {\bibfnamefont {T.}~\bibnamefont {Sugi}}, \bibinfo {author} {\bibfnamefont {F.}~\bibnamefont {Sugihara}}, \bibinfo {author} {\bibfnamefont {K.}~\bibnamefont {Ikeda}}, \bibinfo {author} {\bibfnamefont {H.}~\bibnamefont {Sumiya}}, \bibinfo {author} {\bibfnamefont {S.}~\bibnamefont {Tsuji}}, \bibinfo {author} {\bibfnamefont {I.}~\bibnamefont {Mori}}, \bibinfo {author} {\bibfnamefont {H.}~\bibnamefont {Tochio}}, \bibinfo {author} {\bibfnamefont {Y.}~\bibnamefont {Harada}},\ and\ \bibinfo {author} {\bibfnamefont {M.}~\bibnamefont {Shirakawa}},\ }\href {https://doi.org/10.1021/nl302979d} {\bibfield  {journal} {\bibinfo  {journal} {Nano Letters}\ }\textbf {\bibinfo {volume} {12}},\ \bibinfo {pages} {5726} (\bibinfo {year} {2012})}\BibitemShut {NoStop}%
\bibitem [{\citenamefont {Hui}\ \emph {et~al.}(2014)\citenamefont {Hui}, \citenamefont {Su}, \citenamefont {Chen}, \citenamefont {Chen}, \citenamefont {Liu},\ and\ \citenamefont {Chang}}]{Hui2014wide}%
  \BibitemOpen
  \bibfield  {author} {\bibinfo {author} {\bibfnamefont {Y.~Y.}\ \bibnamefont {Hui}}, \bibinfo {author} {\bibfnamefont {L.-J.}\ \bibnamefont {Su}}, \bibinfo {author} {\bibfnamefont {O.~Y.}\ \bibnamefont {Chen}}, \bibinfo {author} {\bibfnamefont {Y.-T.}\ \bibnamefont {Chen}}, \bibinfo {author} {\bibfnamefont {T.-M.}\ \bibnamefont {Liu}},\ and\ \bibinfo {author} {\bibfnamefont {H.-C.}\ \bibnamefont {Chang}},\ }\href {https://doi.org/10.1038/srep05574} {\bibfield  {journal} {\bibinfo  {journal} {Scientific Reports}\ }\textbf {\bibinfo {volume} {4}},\ \bibinfo {pages} {5574} (\bibinfo {year} {2014})}\BibitemShut {NoStop}%
\bibitem [{\citenamefont {Icha}\ \emph {et~al.}(2017)\citenamefont {Icha}, \citenamefont {Weber}, \citenamefont {Waters},\ and\ \citenamefont {Norden}}]{Icha2017phototoxicity}%
  \BibitemOpen
  \bibfield  {author} {\bibinfo {author} {\bibfnamefont {J.}~\bibnamefont {Icha}}, \bibinfo {author} {\bibfnamefont {M.}~\bibnamefont {Weber}}, \bibinfo {author} {\bibfnamefont {J.~C.}\ \bibnamefont {Waters}},\ and\ \bibinfo {author} {\bibfnamefont {C.}~\bibnamefont {Norden}},\ }\href {https://doi.org/10.1002/bies.201700003} {\bibfield  {journal} {\bibinfo  {journal} {BioEssays}\ }\textbf {\bibinfo {volume} {39}},\ \bibinfo {pages} {1700003} (\bibinfo {year} {2017})}\BibitemShut {NoStop}%
\bibitem [{\citenamefont {Laissue}\ \emph {et~al.}(2017)\citenamefont {Laissue}, \citenamefont {Alghamdi}, \citenamefont {Tomancak}, \citenamefont {Reynaud},\ and\ \citenamefont {Shroff}}]{Laissue2017phototoxicity}%
  \BibitemOpen
  \bibfield  {author} {\bibinfo {author} {\bibfnamefont {P.~P.}\ \bibnamefont {Laissue}}, \bibinfo {author} {\bibfnamefont {R.~A.}\ \bibnamefont {Alghamdi}}, \bibinfo {author} {\bibfnamefont {P.}~\bibnamefont {Tomancak}}, \bibinfo {author} {\bibfnamefont {E.~G.}\ \bibnamefont {Reynaud}},\ and\ \bibinfo {author} {\bibfnamefont {H.}~\bibnamefont {Shroff}},\ }\href {https://doi.org/10.1038/nmeth.4344} {\bibfield  {journal} {\bibinfo  {journal} {Nature Methods}\ }\textbf {\bibinfo {volume} {14}},\ \bibinfo {pages} {657} (\bibinfo {year} {2017})}\BibitemShut {NoStop}%
\bibitem [{\citenamefont {Reineck}\ \emph {et~al.}(2017)\citenamefont {Reineck}, \citenamefont {Capelli}, \citenamefont {Lau}, \citenamefont {Jeske}, \citenamefont {Field}, \citenamefont {Ohshima}, \citenamefont {Greentree},\ and\ \citenamefont {Gibson}}]{Reineck2017Bright}%
  \BibitemOpen
  \bibfield  {author} {\bibinfo {author} {\bibfnamefont {P.}~\bibnamefont {Reineck}}, \bibinfo {author} {\bibfnamefont {M.}~\bibnamefont {Capelli}}, \bibinfo {author} {\bibfnamefont {D.~W.~M.}\ \bibnamefont {Lau}}, \bibinfo {author} {\bibfnamefont {J.}~\bibnamefont {Jeske}}, \bibinfo {author} {\bibfnamefont {M.~R.}\ \bibnamefont {Field}}, \bibinfo {author} {\bibfnamefont {T.}~\bibnamefont {Ohshima}}, \bibinfo {author} {\bibfnamefont {A.~D.}\ \bibnamefont {Greentree}},\ and\ \bibinfo {author} {\bibfnamefont {B.~C.}\ \bibnamefont {Gibson}},\ }\href {https://doi.org/10.1039/C6NR07834F} {\bibfield  {journal} {\bibinfo  {journal} {Nanoscale}\ }\textbf {\bibinfo {volume} {9}},\ \bibinfo {pages} {497} (\bibinfo {year} {2017})}\BibitemShut {NoStop}%
\bibitem [{\citenamefont {Oshimi}\ \emph {et~al.}(2024)\citenamefont {Oshimi}, \citenamefont {Ishiwata}, \citenamefont {Nakashima}, \citenamefont {Mandic}, \citenamefont {Kobayashi}, \citenamefont {Teramoto}, \citenamefont {Tsuji}, \citenamefont {Nishibayashi}, \citenamefont {Shikano}, \citenamefont {An},\ and\ \citenamefont {Fujiwara}}]{Oshimi2024bright}%
  \BibitemOpen
  \bibfield  {author} {\bibinfo {author} {\bibfnamefont {K.}~\bibnamefont {Oshimi}}, \bibinfo {author} {\bibfnamefont {H.}~\bibnamefont {Ishiwata}}, \bibinfo {author} {\bibfnamefont {H.}~\bibnamefont {Nakashima}}, \bibinfo {author} {\bibfnamefont {S.}~\bibnamefont {Mandic}}, \bibinfo {author} {\bibfnamefont {H.}~\bibnamefont {Kobayashi}}, \bibinfo {author} {\bibfnamefont {M.}~\bibnamefont {Teramoto}}, \bibinfo {author} {\bibfnamefont {H.}~\bibnamefont {Tsuji}}, \bibinfo {author} {\bibfnamefont {Y.}~\bibnamefont {Nishibayashi}}, \bibinfo {author} {\bibfnamefont {Y.}~\bibnamefont {Shikano}}, \bibinfo {author} {\bibfnamefont {T.}~\bibnamefont {An}},\ and\ \bibinfo {author} {\bibfnamefont {M.}~\bibnamefont {Fujiwara}},\ }\href {https://doi.org/10.1021/acsnano.4c03424} {\bibfield  {journal} {\bibinfo  {journal} {ACS Nano}\ }\textbf {\bibinfo {volume} {18}},\ \bibinfo {pages} {35202} (\bibinfo {year} {2024})}\BibitemShut {NoStop}%
\bibitem [{\citenamefont {Hopper}\ \emph {et~al.}(2018)\citenamefont {Hopper}, \citenamefont {Shulevitz},\ and\ \citenamefont {Bassett}}]{Hopper2018spinreadout}%
  \BibitemOpen
  \bibfield  {author} {\bibinfo {author} {\bibfnamefont {D.~A.}\ \bibnamefont {Hopper}}, \bibinfo {author} {\bibfnamefont {H.~J.}\ \bibnamefont {Shulevitz}},\ and\ \bibinfo {author} {\bibfnamefont {L.~C.}\ \bibnamefont {Bassett}},\ }\href {https://doi.org/10.3390/mi9090437} {\bibfield  {journal} {\bibinfo  {journal} {Micromachines}\ }\textbf {\bibinfo {volume} {9}},\ \bibinfo {pages} {437} (\bibinfo {year} {2018})}\BibitemShut {NoStop}%
\bibitem [{\citenamefont {Pasienski}\ and\ \citenamefont {DeMarco}(2008)}]{Pasienski2008holographic}%
  \BibitemOpen
  \bibfield  {author} {\bibinfo {author} {\bibfnamefont {M.}~\bibnamefont {Pasienski}}\ and\ \bibinfo {author} {\bibfnamefont {B.}~\bibnamefont {DeMarco}},\ }\href {https://doi.org/10.1364/OE.16.002176} {\bibfield  {journal} {\bibinfo  {journal} {Optics Express}\ }\textbf {\bibinfo {volume} {16}},\ \bibinfo {pages} {2176} (\bibinfo {year} {2008})}\BibitemShut {NoStop}%
\bibitem [{\citenamefont {Gaunt}\ and\ \citenamefont {Hadzibabic}(2012)}]{Gaunt2012holography}%
  \BibitemOpen
  \bibfield  {author} {\bibinfo {author} {\bibfnamefont {A.~L.}\ \bibnamefont {Gaunt}}\ and\ \bibinfo {author} {\bibfnamefont {Z.}~\bibnamefont {Hadzibabic}},\ }\href {https://doi.org/10.1038/srep00721} {\bibfield  {journal} {\bibinfo  {journal} {Scientific Reports}\ }\textbf {\bibinfo {volume} {2}},\ \bibinfo {pages} {721} (\bibinfo {year} {2012})}\BibitemShut {NoStop}%
\bibitem [{\citenamefont {Nogrette}\ \emph {et~al.}(2014)\citenamefont {Nogrette}, \citenamefont {Labuhn}, \citenamefont {Ravets}, \citenamefont {Barredo}, \citenamefont {B{\'e}guin}, \citenamefont {Vernier}, \citenamefont {Lahaye},\ and\ \citenamefont {Browaeys}}]{Nogrette2014holographic}%
  \BibitemOpen
  \bibfield  {author} {\bibinfo {author} {\bibfnamefont {F.}~\bibnamefont {Nogrette}}, \bibinfo {author} {\bibfnamefont {H.}~\bibnamefont {Labuhn}}, \bibinfo {author} {\bibfnamefont {S.}~\bibnamefont {Ravets}}, \bibinfo {author} {\bibfnamefont {D.}~\bibnamefont {Barredo}}, \bibinfo {author} {\bibfnamefont {L.}~\bibnamefont {B{\'e}guin}}, \bibinfo {author} {\bibfnamefont {A.}~\bibnamefont {Vernier}}, \bibinfo {author} {\bibfnamefont {T.}~\bibnamefont {Lahaye}},\ and\ \bibinfo {author} {\bibfnamefont {A.}~\bibnamefont {Browaeys}},\ }\href {https://doi.org/10.1103/PhysRevX.4.021034} {\bibfield  {journal} {\bibinfo  {journal} {Physical Review X}\ }\textbf {\bibinfo {volume} {4}},\ \bibinfo {pages} {021034} (\bibinfo {year} {2014})}\BibitemShut {NoStop}%
\bibitem [{\citenamefont {Cai}\ \emph {et~al.}(2021)\citenamefont {Cai}, \citenamefont {Guo}, \citenamefont {Shi}, \citenamefont {Li}, \citenamefont {Wang}, \citenamefont {Ji}, \citenamefont {Wang},\ and\ \citenamefont {Du}}]{Cai2021parallel}%
  \BibitemOpen
  \bibfield  {author} {\bibinfo {author} {\bibfnamefont {M.}~\bibnamefont {Cai}}, \bibinfo {author} {\bibfnamefont {Z.}~\bibnamefont {Guo}}, \bibinfo {author} {\bibfnamefont {F.}~\bibnamefont {Shi}}, \bibinfo {author} {\bibfnamefont {C.}~\bibnamefont {Li}}, \bibinfo {author} {\bibfnamefont {M.}~\bibnamefont {Wang}}, \bibinfo {author} {\bibfnamefont {W.}~\bibnamefont {Ji}}, \bibinfo {author} {\bibfnamefont {P.}~\bibnamefont {Wang}},\ and\ \bibinfo {author} {\bibfnamefont {J.}~\bibnamefont {Du}},\ }\href {https://doi.org/10.1063/5.0039110} {\bibfield  {journal} {\bibinfo  {journal} {Review of Scientific Instruments}\ }\textbf {\bibinfo {volume} {92}},\ \bibinfo {pages} {045107} (\bibinfo {year} {2021})}\BibitemShut {NoStop}%
\bibitem [{\citenamefont {Cai}\ \emph {et~al.}(2023)\citenamefont {Cai}, \citenamefont {Guo}, \citenamefont {Li}, \citenamefont {Li}, \citenamefont {Shi}, \citenamefont {Wang}, \citenamefont {Li}, \citenamefont {Wang},\ and\ \citenamefont {Du}}]{Cai2023highthroughput}%
  \BibitemOpen
  \bibfield  {author} {\bibinfo {author} {\bibfnamefont {M.}~\bibnamefont {Cai}}, \bibinfo {author} {\bibfnamefont {Z.}~\bibnamefont {Guo}}, \bibinfo {author} {\bibfnamefont {C.}~\bibnamefont {Li}}, \bibinfo {author} {\bibfnamefont {S.}~\bibnamefont {Li}}, \bibinfo {author} {\bibfnamefont {F.}~\bibnamefont {Shi}}, \bibinfo {author} {\bibfnamefont {M.}~\bibnamefont {Wang}}, \bibinfo {author} {\bibfnamefont {H.}~\bibnamefont {Li}}, \bibinfo {author} {\bibfnamefont {P.}~\bibnamefont {Wang}},\ and\ \bibinfo {author} {\bibfnamefont {J.}~\bibnamefont {Du}},\ }\href {https://doi.org/10.1002/qute.202200175} {\bibfield  {journal} {\bibinfo  {journal} {Advanced Quantum Technologies}\ }\textbf {\bibinfo {volume} {6}},\ \bibinfo {pages} {2200175} (\bibinfo {year} {2023})}\BibitemShut {NoStop}%
\bibitem [{\citenamefont {Tamura}\ \emph {et~al.}(2016)\citenamefont {Tamura}, \citenamefont {Unakami}, \citenamefont {He}, \citenamefont {Miyamoto},\ and\ \citenamefont {Nakagawa}}]{Tamura2016highly}%
  \BibitemOpen
  \bibfield  {author} {\bibinfo {author} {\bibfnamefont {H.}~\bibnamefont {Tamura}}, \bibinfo {author} {\bibfnamefont {T.}~\bibnamefont {Unakami}}, \bibinfo {author} {\bibfnamefont {J.}~\bibnamefont {He}}, \bibinfo {author} {\bibfnamefont {Y.}~\bibnamefont {Miyamoto}},\ and\ \bibinfo {author} {\bibfnamefont {K.}~\bibnamefont {Nakagawa}},\ }\href {https://doi.org/10.1364/OE.24.008132} {\bibfield  {journal} {\bibinfo  {journal} {Optics Express}\ }\textbf {\bibinfo {volume} {24}},\ \bibinfo {pages} {8132} (\bibinfo {year} {2016})}\BibitemShut {NoStop}%
\bibitem [{\citenamefont {Kim}\ \emph {et~al.}(2019)\citenamefont {Kim}, \citenamefont {Keesling}, \citenamefont {Omran}, \citenamefont {Levine}, \citenamefont {Bernien}, \citenamefont {Greiner}, \citenamefont {Lukin},\ and\ \citenamefont {Englund}}]{Kim2019large}%
  \BibitemOpen
  \bibfield  {author} {\bibinfo {author} {\bibfnamefont {D.}~\bibnamefont {Kim}}, \bibinfo {author} {\bibfnamefont {A.}~\bibnamefont {Keesling}}, \bibinfo {author} {\bibfnamefont {A.}~\bibnamefont {Omran}}, \bibinfo {author} {\bibfnamefont {H.}~\bibnamefont {Levine}}, \bibinfo {author} {\bibfnamefont {H.}~\bibnamefont {Bernien}}, \bibinfo {author} {\bibfnamefont {M.}~\bibnamefont {Greiner}}, \bibinfo {author} {\bibfnamefont {M.~D.}\ \bibnamefont {Lukin}},\ and\ \bibinfo {author} {\bibfnamefont {D.~R.}\ \bibnamefont {Englund}},\ }\href {https://doi.org/10.1364/OL.44.003178} {\bibfield  {journal} {\bibinfo  {journal} {Optics Letters}\ }\textbf {\bibinfo {volume} {44}},\ \bibinfo {pages} {3178} (\bibinfo {year} {2019})}\BibitemShut {NoStop}%
\bibitem [{\citenamefont {Chew}\ \emph {et~al.}(2024)\citenamefont {Chew}, \citenamefont {Poitrinal}, \citenamefont {Tomita}, \citenamefont {Kitade}, \citenamefont {Mauricio}, \citenamefont {Ohmori},\ and\ \citenamefont {de~L{\'e}s{\'e}leuc}}]{Chew2024ultraprecise}%
  \BibitemOpen
  \bibfield  {author} {\bibinfo {author} {\bibfnamefont {Y.~T.}\ \bibnamefont {Chew}}, \bibinfo {author} {\bibfnamefont {M.}~\bibnamefont {Poitrinal}}, \bibinfo {author} {\bibfnamefont {T.}~\bibnamefont {Tomita}}, \bibinfo {author} {\bibfnamefont {S.}~\bibnamefont {Kitade}}, \bibinfo {author} {\bibfnamefont {J.}~\bibnamefont {Mauricio}}, \bibinfo {author} {\bibfnamefont {K.}~\bibnamefont {Ohmori}},\ and\ \bibinfo {author} {\bibfnamefont {S.}~\bibnamefont {de~L{\'e}s{\'e}leuc}},\ }\href {https://doi.org/10.1103/PhysRevA.110.053518} {\bibfield  {journal} {\bibinfo  {journal} {Physical Review A}\ }\textbf {\bibinfo {volume} {110}},\ \bibinfo {pages} {053518} (\bibinfo {year} {2024})}\BibitemShut {NoStop}%
\bibitem [{\citenamefont {Cheng}\ \emph {et~al.}(2025)\citenamefont {Cheng}, \citenamefont {Kazi}, \citenamefont {Rovny}, \citenamefont {Zhang}, \citenamefont {Nassar}, \citenamefont {Thompson},\ and\ \citenamefont {de~Leon}}]{Cheng2025massively}%
  \BibitemOpen
  \bibfield  {author} {\bibinfo {author} {\bibfnamefont {K.-H.}\ \bibnamefont {Cheng}}, \bibinfo {author} {\bibfnamefont {Z.}~\bibnamefont {Kazi}}, \bibinfo {author} {\bibfnamefont {J.}~\bibnamefont {Rovny}}, \bibinfo {author} {\bibfnamefont {B.}~\bibnamefont {Zhang}}, \bibinfo {author} {\bibfnamefont {L.~S.}\ \bibnamefont {Nassar}}, \bibinfo {author} {\bibfnamefont {J.~D.}\ \bibnamefont {Thompson}},\ and\ \bibinfo {author} {\bibfnamefont {N.~P.}\ \bibnamefont {de~Leon}},\ }\href {https://doi.org/10.1103/t8fz-3tzs} {\bibfield  {journal} {\bibinfo  {journal} {Physical Review X}\ }\textbf {\bibinfo {volume} {15}},\ \bibinfo {pages} {031014} (\bibinfo {year} {2025})}\BibitemShut {NoStop}%
\bibitem [{\citenamefont {Cambria}\ \emph {et~al.}(2025)\citenamefont {Cambria}, \citenamefont {Chand}, \citenamefont {Reiter},\ and\ \citenamefont {Kolkowitz}}]{Cambria2025scalable}%
  \BibitemOpen
  \bibfield  {author} {\bibinfo {author} {\bibfnamefont {M.}~\bibnamefont {Cambria}}, \bibinfo {author} {\bibfnamefont {S.}~\bibnamefont {Chand}}, \bibinfo {author} {\bibfnamefont {C.~M.}\ \bibnamefont {Reiter}},\ and\ \bibinfo {author} {\bibfnamefont {S.}~\bibnamefont {Kolkowitz}},\ }\href {https://doi.org/10.1103/jdzq-jbfz} {\bibfield  {journal} {\bibinfo  {journal} {Physical Review X}\ }\textbf {\bibinfo {volume} {15}},\ \bibinfo {pages} {031015} (\bibinfo {year} {2025})}\BibitemShut {NoStop}%
\bibitem [{\citenamefont {Schroff}\ \emph {et~al.}(2024)\citenamefont {Schroff}, \citenamefont {Haller}, \citenamefont {Kuhr},\ and\ \citenamefont {La~Rooij}}]{Schroff2024rapid}%
  \BibitemOpen
  \bibfield  {author} {\bibinfo {author} {\bibfnamefont {P.}~\bibnamefont {Schroff}}, \bibinfo {author} {\bibfnamefont {E.}~\bibnamefont {Haller}}, \bibinfo {author} {\bibfnamefont {S.}~\bibnamefont {Kuhr}},\ and\ \bibinfo {author} {\bibfnamefont {A.}~\bibnamefont {La~Rooij}},\ }\href {https://doi.org/10.1364/OE.539548} {\bibfield  {journal} {\bibinfo  {journal} {Optics Express}\ }\textbf {\bibinfo {volume} {32}},\ \bibinfo {pages} {48957} (\bibinfo {year} {2024})}\BibitemShut {NoStop}%
\bibitem [{\citenamefont {Di~Leonardo}\ \emph {et~al.}(2007)\citenamefont {Di~Leonardo}, \citenamefont {Ianni},\ and\ \citenamefont {Ruocco}}]{DiLeonardo2007computer}%
  \BibitemOpen
  \bibfield  {author} {\bibinfo {author} {\bibfnamefont {R.}~\bibnamefont {Di~Leonardo}}, \bibinfo {author} {\bibfnamefont {F.}~\bibnamefont {Ianni}},\ and\ \bibinfo {author} {\bibfnamefont {G.}~\bibnamefont {Ruocco}},\ }\href {https://doi.org/10.1364/OE.15.001913} {\bibfield  {journal} {\bibinfo  {journal} {Optics Express}\ }\textbf {\bibinfo {volume} {15}},\ \bibinfo {pages} {1913} (\bibinfo {year} {2007})}\BibitemShut {NoStop}%
\bibitem [{\citenamefont {Ronzitti}\ \emph {et~al.}(2012)\citenamefont {Ronzitti}, \citenamefont {Guillon}, \citenamefont {de~Sars},\ and\ \citenamefont {Emiliani}}]{Ronzitti2012slm}%
  \BibitemOpen
  \bibfield  {author} {\bibinfo {author} {\bibfnamefont {E.}~\bibnamefont {Ronzitti}}, \bibinfo {author} {\bibfnamefont {M.}~\bibnamefont {Guillon}}, \bibinfo {author} {\bibfnamefont {V.}~\bibnamefont {de~Sars}},\ and\ \bibinfo {author} {\bibfnamefont {V.}~\bibnamefont {Emiliani}},\ }\href {https://doi.org/10.1364/OE.20.017843} {\bibfield  {journal} {\bibinfo  {journal} {Optics Express}\ }\textbf {\bibinfo {volume} {20}},\ \bibinfo {pages} {17843} (\bibinfo {year} {2012})}\BibitemShut {NoStop}%
\bibitem [{\citenamefont {Catal{\`a}-Castro}\ and\ \citenamefont {Mart{\'i}n-Badosa}(2021)}]{CatalaCastro2021positioning}%
  \BibitemOpen
  \bibfield  {author} {\bibinfo {author} {\bibfnamefont {F.}~\bibnamefont {Catal{\`a}-Castro}}\ and\ \bibinfo {author} {\bibfnamefont {E.}~\bibnamefont {Mart{\'i}n-Badosa}},\ }\href {https://doi.org/10.3390/mi12050559} {\bibfield  {journal} {\bibinfo  {journal} {Micromachines}\ }\textbf {\bibinfo {volume} {12}},\ \bibinfo {pages} {559} (\bibinfo {year} {2021})}\BibitemShut {NoStop}%
\end{thebibliography}%
\end{document}